\documentclass[aps,prb,twocolumn,showpacs]{revtex4-1}

\usepackage{graphicx}
\usepackage{multirow}
\usepackage{fixltx2e}

\usepackage{amsmath}
\usepackage[normalem]{ulem}
\usepackage{color}

\begin{document}

\title{Borophane as substrate for adsorption of $^4$He: A journey across dimensionality.}

\author{Stefania De Palo}
\affiliation{CNR-IOM Democritos, via Bonomea, 265 - 34136 Trieste, Italy}
\affiliation{SISSA (International School for Advanced Studies), via Bonomea 265, 34136 Trieste, Italy}
\author{Saverio Moroni}
\affiliation{CNR-IOM Democritos, via Bonomea, 265 - 34136 Trieste, Italy}
\affiliation{SISSA (International School for Advanced Studies), via Bonomea 265, 34136 Trieste, Italy}
\author{Francesco Ancilotto}
\affiliation{Dipartimento di Fisica e Astronomia "G.Galilei" and CNISM, Universit\`a di Padova, Via Marzolo 8 I-35131 Padova Italy}
\affiliation{CNR-IOM Democritos, via Bonomea, 265 - 34136 Trieste, Italy}
\author{Pier Luigi Silvestrelli}
\affiliation{Dipartimento di Fisica e Astronomia "G.Galilei" and CNISM, Universit\`a di Padova, Via Marzolo 8 I-35131 Padova Italy}
\affiliation{CNR-IOM Democritos, via Bonomea, 265 - 34136 Trieste, Italy}
\author{Luciano Reatto}
\affiliation{Dipartimento di Fisica, Universit\`a degli Studi di Milano, via Celoria 16, 20133 Milano, Italy}

\begin{abstract}
In search of substrates for adsorption of He atoms allowing for novel 
quantum phases in restricted geometry, we study the case of borophane,
a hydrogenated compound of borophene. 
We consider two allotropes of borophane, $\alpha' -4$H and Rect–2H. 
With a suitable Density Functional Theory we characterize the adsorption 
potential of a He atom on such crystalline substrates finding its corrugation, 
the preferential adsorption sites, and the energy barrier between
sites. In the case of $\alpha' -4$H borophane the adsorption potential 
has some similarity to that of graphite but with much larger 
energy barriers between adsorption sites so that the first
adsorbed layer of $^4$He should be localized in a 
registered triangular crystal similar to the 
$\sqrt{3}\times\sqrt{3}{\rm R}30^{\circ}$ phase on graphite. 
Rect-2H borophane appears more interesting due to the 
presence of ridges in the adsorption potential with modest energy 
barriers in one direction of the basal plane and much higher barrier in
the orthogonal direction, thus forming channels for motion of the adsorbed 
atoms. Using Path Integral Monte Carlo (PIMC) simulations 
we find that in the first adsorbed layer the $^4$He atoms 
are rather delocalized along a channel with no exchanges between channels. 
This strong anisotropy is present also in the first few additional 
adsorption layers with presence of ordered and of disordered 
regions. In the second, fourth and fifth layers we find superfluidity on 
the length scale of the simulated systems. 
In the second layer, the superfluidity is one–dimensional along the grooves. 
In a narrow low-coverage region of the fourth layer we observe an intriguing
state with one unidirectional structure across the grooves which supports
a tiny superfluid signal. 
In the fifth layer we find a two-dimensional superfluid, with a crossover
from strongly {\em anisotropic} at low coverage to {\em isotropic}
at layer completion. Starting from the sixth layer, the adsorbed
$^4$He film evolves towards a three–dimensional superfluid. 
Our main prediction is that adsorption of $^4$He on Rect-2H borophane 
will allow to probe 1D superfluidity in the second and possibly the fourth
layer, the evolution from a 2D anisotropic superfluid to an isotropic one in 
the fifth layer, and eventually the onset of 3D superfluidity for higher 
coverages. 

\end{abstract}

\maketitle

\section{Introduction}
The experimental realization of borophene (a two-dimensional material made of 
boron atoms) has triggered a great deal
of interest both theoretically and 
experimentally
due to its peculiar 
mechanical, electronic, and optical properties that make it 
attractive in a number of different areas of science and technology
such as energy storage, flexible electronics or photonics and  
optoelectronics.
\cite{review}
However, free-standing borophene sheets were predicted to be dynamically unstable,\cite{mannix}
which hinders the practical use of this material. 
Various methods were proposed to improve the stability of borophene such 
as adsorption on suitable metal substrates or surface functionalization.
Recently, an ordered and stable heterostructure, named {\it borophane}, has 
been synthesized by hydrogenating borophene with atomic hydrogen.\cite{qli,hou}

Here we address borophane as a substrate for adsorption 
studies, specifically we characterize the adsorption of $^4$He atoms. 
Adsorbed phases of the He isotopes $^4$He and $^3$He have been studied 
on a number of substrates for many decades, to explore the
effect of reduced dimensionality on quantum properties such as 
superfluidity in $^4$He\cite{crowell_1996} or the role of statistics,
Bose versus Fermi.\cite{greywall_1990_1991} 
On disordered and strongly binding substrates like vycor or metals, 
with the exception of some alkali metals, in the first few adsorbed 
layers the He atoms are localized and disordered. When an additional 
layer of $^4$He atoms is added above such ``dead layers`''
one finds an almost uniform fluid which at low temperature 
is superfluid.\cite{crowell_1997}
In the case of the highly ordered substrate graphite the first 
adsorbed layer at low temperature is dominated by an ordered state 
in which the He atoms are localized at a subset of the adsorption 
sites that are provided by the substrate.\cite{bruch_2007} 
Theory\cite{reatto_2013} predicts that the same behavior occurs 
in the case of adsorption on graphene. 
A novel behavior has been  
predicted\cite{moroni21} instead in the case of the adsorption 
of $^4$He on fluorographene (GF), a sheet of graphene in which a 
fluorine atom is chemically bound to each carbon atom. 
Fluorine atoms modify the adsorption sites, changing their density, 
and 
corrugation of the adsorption potential. The main 
theoretical result is that the ground state of a monolayer 
of $^4$He atoms is a spatially modulated superfluid. The density 
inhomogeneity is very large, so one has at hand a superfluid that 
is inhomogeneous on an atomic scale, a regime quite different from that 
found on other adsorption substrates for which the modulation at most 
is very weak.  
For example,
roton excitations of $^4$He on GF have 
been predicted to yield a large anisotropic dispersion 
relation.\cite{nava_2013}

The availability of new ordered 
two-dimensional systems 
like borophene and its derivatives, such as borophane, 
motivates a study of the adsorption of He to expand our
understanding of strongly 
interacting quantum particles in 
modulating external potential. 
In the field of quantum particles in a modulating external potential,
remarkable experiments have been performed with cold atoms in optical lattices
\cite{yukalov_2009} but typically in these systems the period of the 
external potential
is orders of magnitude larger than the range of the interatomic potential
or of the scattering length.
Quite differently, in the case of adsorbed phases of He one can explore 
the behavior of systems in which the ranges 
of the interatomic potential and of the average interparticle distance 
are comparable to the period of the modulating potential. 

Borophene is a strongly reacting system, so
an experimental study of adsorption on borophene is problematic. 
In contrast
borophane is quite stable\cite{hou,qli} so that an adsorption 
study should be experimentally feasible.  
Borophene has a high degree of polymorphism, and the same 
holds for borophane.
Here we investigate theoretically, from first principles, the adsorption 
of $^4$He on two polymorphs of borophane, which 
are termed $\alpha '-4$H\cite{hou} 
and Rect-2H,\cite{qli} respectively. The
former is very stable and also produced without any metal substrate.\cite{hou}
The second polymorph has been grown on a metallic substrate.


The adsorption potentials, characterizing the interaction of a $^4$He atom 
with these two polymorphs of borophane, are computed 
using ab initio calculations, in the framework of the Density Functional Theory (DFT)
with a suitable functional theory, 
capable of properly including van der Waals 
(vdW) interactions. 
We find that the adsorption energies of $^4$He on these two polymorphs of
borophane are almost the same and rather close to 
those on graphene/graphite.

In the case of $\alpha '-4$H borophane the adsorption sites
have the same symmetry as in the case of graphene/graphite, i.e.,
these sites form a triangular lattice and its lattice spacing
is close to the second neighbor intersite distance in graphene/graphite.
According to our calculations, the corrugation of the 
adsorption potential is 
very large, the intersite barrier is more than 80\% larger
than in the case of graphene/graphite. The curvature of the adsorption potential in the $z$
direction at an adsorption site in $\alpha '-4$H borophane is similar to that for 
graphene/graphite, so that the zero point motion for localization perpendicular to the basal plane is
similar in the two cases. 
We conclude
that the ground state of a monolayer of $^4$He on $\alpha '-4$H should be an ordered state 
with 
atoms localized at adsorption sites with triangular symmetry 
as on graphene/graphite, 
that is the famous $\sqrt{3} \times \sqrt{3}$R$30^{\circ}$ phase
in which $^4$He atoms occupy second
neighbor sites of the triangular lattice of adsorption sites.
 
More interesting is the case of Rect-2H borophane. This polymorph is not symmetric in the 
direction perpendicular to the basal plane, so 
the adsorption potential is different above or below the basal plane. On both sides, the most favored adsorption sites form 
a rectangular lattice with the sides in the ratio 1.69/1. 
In addition, the ratio between the potential 
energy barriers between the adsorption sites along the $y$ direction and along the $x$ 
direction is 2.8 above the basal plane and 2.5 below the basal plane. Therefore, the adsorption potential is highly anisotropic in the $x$-$y$ 
basal plane. 

Given such features of the adsorption potential of $^4$He on Rect-2H borophane we anticipate
that the behavior of the adsorbed $^4$He should be quite distinct from that of other known
substrates. To verify this,
we have characterized the absorption of the first several layers of $^4$He atoms on the 
upper surface of Rect--2H borophane 
by performing path integral Monte Carlo (PIMC) simulations in the
grand canonical ensemble.
As the first layer of $^4$He grows the atoms are found in a 
disordered state, not superfluid at the temperatures of our computations, and this state
becomes ordered in register with the substrate at completion of the first layer.
At higher coverage beyond the first layer, depending on the chemical potential, 
the system can be disordered, ordered, or in a mixed state, 
partially ordered and partially disordered depending on the 
distance of the atoms from the basal plane. 
The strong anisotropy of the adsorption potential in the $(x,y)$ plane has a strong 
effect on the properties of the adsorbed atoms: these form rows with a very low 
probability of exchange between atoms in different rows, whereas exchanges between atoms 
in the same row take place much more easily. In fact, we find one--dimensional (1D) 
superfluidity on the length scale of the simulated system at low temperature and for 
certain ranges of the coverage. However, the system cannot be considered as an ensemble of 
independent rows because, even in absence of exchanges, neighboring rows turn out to be 
correlated due to the interparticle interaction. 
This 1D superfluidity disappears for larger coverage in favor of an ordered or partially ordered
phase. In the fifth layer, a 2D anisotropic superfluidity
emerges, which crosses over to isotropic at layer 
completion, and evolves toward bulk superfluidity at higher
coverages. This complex evolution of the adsorbed $^4$He 
appears to be a novel regime not yet observed in the adsorbed
phases.

The paper is organized as follows. In Sect. II the adsorption potential is derived
and its main features are characterized. In Sect. III the Quantum Monte Carlo methods used in our calculations
are briefly introduced and the results for the adsorption of $^4$He on borophane 
are presented. A summary and our conclusions are given in Sect. IV.

\section{Adsorption potential}
An essential ingredient of quantum simulations
of He adsorption on surfaces
is an accurate description of the
interaction between 
a He atom and the substrates.
Similarly to what was done in a previous work,\cite{moroni21} where 
submonolayer $^4$He on fluorographene, hexagonal boron nitride, and graphene were studied,
we derived from first--principles calculations 
the interaction of He atoms with the borophane substrate, 
using state-of-the-art DFT functionals
specifically designed to accurately describe the weak vdW interactions,
thus providing an accurate description
of the interaction of He atoms with the substrate.
Recent applications of vdW-corrected DFT schemes 
containing corrections aimed at reproducing vdW forces to the
problem of atoms/molecules-surface interactions
have proven the accuracy of such methods
in the calculation of both adsorption distances and adsorption
energies
as well as the high degree of reliability
across a wide range of adsorbates.

We have computed the He atom adsorption energies on different surface sites
and the potential energy corrugations along the plane,
which are the most crucial ingredients for accurate quantum simulations
of the adsorption of $^4$He.
Our calculations have been performed
with the Quantum-ESPRESSO {\it ab initio} package.\cite{ESPRESSO}
A single He atom per supercell is considered and we model the substrates
adopting periodically repeated orthorhombic supercells.
The lattice constant has been optimized by requiring that it corresponds to the 
minimum-energy equilibrium state of the substrates. 
Repeated slabs were separated along the direction orthogonal to the 
surface by a vacuum region
of about 24\AA ~to avoid significant spurious interactions due to periodic
replicas. The Brillouin Zone has been sampled using a
$2\times 2\times 1$ $k$-point mesh.
Electron-ion interactions were described using ultrasoft
pseudopotentials and the wavefunctions were expanded in a plane-wave basis 
set with an energy cutoff of 50 Ry.

The calculations have been performed by adopting the  
rVV10 DFT functional\cite{Sabatini} (this is the revised and more efficient 
version of the original VV10 scheme\cite{Vydrov}), where
vdW effects are included by introducing an explicitly non--local correlation 
functional. 
rVV10 has been found to perform well in many systems and processes 
where vdW effects are relevant, including several adsorption 
processes.\cite{Sabatini,psil15,psil16}

\begin{figure}[h]
\includegraphics[width=10cm]{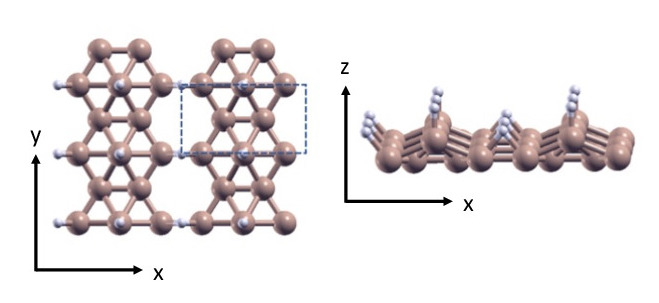}
\caption{Top and side views of the supercell for the
Rect-2H borophane structure.
The unit cell is denoted by dashed lines.
The sides of the unit cell are
$a=5.115$\AA\ ~and $b=2.855$\AA.}
\label{fig2}
\end{figure}

\begin{figure}[h]
\includegraphics[width=5.5cm]{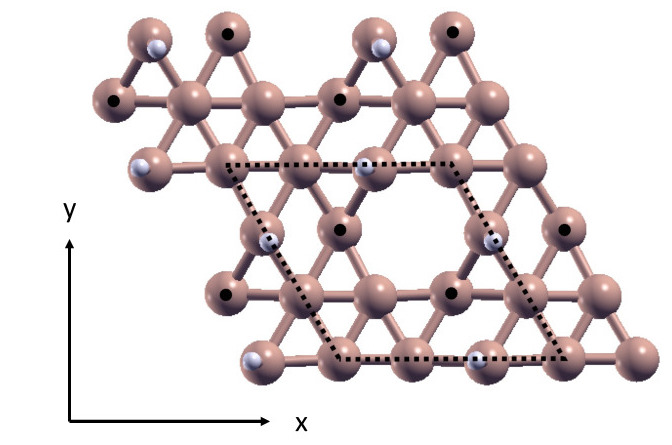}
\includegraphics[width=4.5cm]{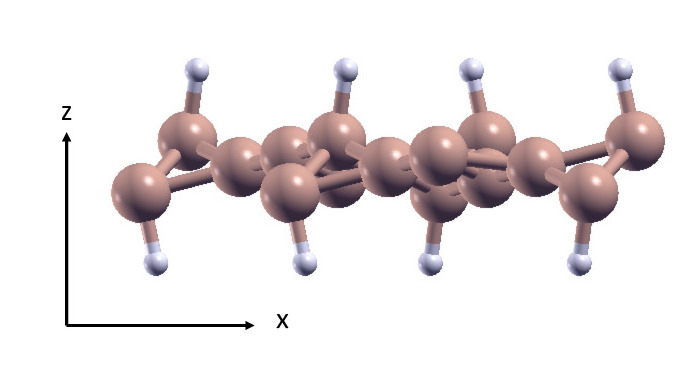}
\caption{Top and side views of the supercell for the
$\alpha '-4$H borophane structure.
The unit cell is denoted by dashed lines.
Balls with black dots denote B atoms bonded to an underlying H atom.
The side of the unit cell is $a=5.061$\AA.}
\label{fig1}
\end{figure}

\begin{figure}[h]
\includegraphics[width=8cm]{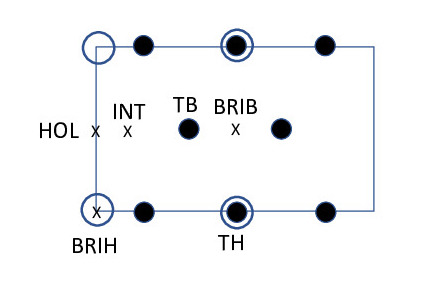}
\caption{Sites chosen for the calculation of the He-borophane potentials,
considering the Rect-2H borophane structure. 
The open and filled circles represent Hydrogen and Boron, respectively;
the crosses are the ($x,y$) coordinates of the He atom. 
}
\label{fig3}
\end{figure}

\begin{figure}[h]
{\vskip -3.5cm}
\includegraphics[width=8cm]{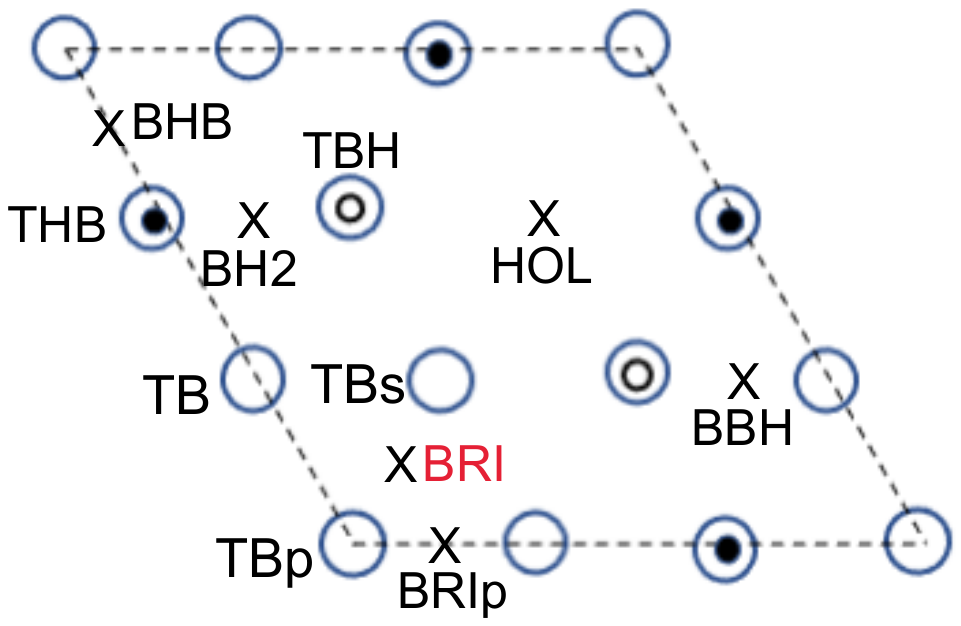}
{\vskip -3.5cm}
\caption{Sites chosen for the calculation of the He-borophane potentials,
considering the $\alpha '-4$H borophane structure.
The open circles represent Boron atoms, the smaller open (filled) 
circles Hydrogen atoms below (above) Boron, and 
the crosses are the ($x,y$) coordinates of the He atom.}
\label{fig1bis}
\end{figure}

The structure of the two polymorphs Rect--2H and $\alpha '-4$H 
is shown in Figs. (\ref{fig2}) and (\ref{fig1}), respectively. 
These configurations can be considered as derived from the
$\beta_{12}$ and $\alpha$ borophene structures, respectively.\cite{Gozar}
We computed the He-substrate interaction 
for a selected set of nonequivalent 
sites in the primitive surface unit cell (see
Figs. (\ref{fig3}) and (\ref{fig1bis}) ). 

For Rect--2H 
we chose the following points:
HOL (hollow site), TB (site on top of B atom), TH (site on top of H 
atom), BRIH (bridge site on top of a H atom between two B atoms);
INT (intermediate site between HOL and TB); 
BRIB (bridge site between two B atoms).
The chosen sites are shown in Fig.(\ref{fig3}); we have considered 
the adsorption of He on both the upper and lower surfaces.

For each chosen point of $^4$He we computed the energy as a function of the $z$ coordinate
finding the lowest value.
Our most relevant numerical results for the adsorption of 
He on the Rect-2H borophane 
substrate are summarized in Tables I and II. 

As can be seen, the most stable adsorption site (for both the upper and lower surfaces)
for He on Rect-2H borophane is the HOL site. 
The least favored sites, characterized by the highest-energy,
are instead the TH site (for the upper surface) and the BRIB site (for the 
lower surface).
The minimum intersite barriers $\Delta_x$ and $\Delta_y$, reported in Table II,
are given by the energy barriers between two 
adsorption sites for displacements constrained to the $x$ and $y$ 
directions, respectively, while optimizing the $z$ coordinate.
They are given by the binding energy differences between the BRIB and HOL sites
along $x$ and the BRIH and HOL sites along $y$. Clearly $\Delta_x$ is much larger
than the minimum energy barrier $\Delta_y$ and for the upper surface $\Delta_y$=47 K is
equal to the value for a graphene substrate.\cite{moroni21}

\begin{table}
\caption{Binding energy, $E_b$ (from the most favored to the least favored configuration), 
for He on the Rect-2H borophane
substrate (considering both upper and lower surface), relative to the
sites shown in Fig.(\ref{fig3}), in parenthesis values obtained
by replacing the rVV10 DFT functional with DFT-D2.}
\begin{tabular}{|l|c|c|}
\hline\noalign{\smallskip}
 site       &  $E_b$(K) (upper surface) & $E_b$(K) (lower surface)  \\ 
\hline
 HOL        &  -268 (-344)              &  -245  \\
 INT        &  -244 (-325)              &  -224  \\
 BRIH       &  -221 (-291)              &  -212  \\
 TB         &  -142 (-175)              &  -170  \\
 BRIB       &  -133 (-162)              &  -163  \\
 TH         &  -123 (-136)              &  -180  \\
\noalign{\smallskip}\hline
\end{tabular}
\label{table-rect}
\end{table}


\begin{table}
\caption{Binding energy in the most-favored configuration
for He on the Rect-2H borophane structure, $E_b$,
distance of the He atom from the substrate, $d$, 
and minimum intersite barriers (see text) along the $x$ and $y$ directions,
$\Delta_{x}$ and $\Delta_{y}$. 
}
\begin{tabular}{|c|c|c|c|c|}
\hline\noalign{\smallskip}
                &  $E_b$(K) & $d(\rm{\AA})$ &$\Delta_x$(K) &$\Delta_y$(K) \\ 
\hline
 upper surface  &   -268    &  3.4          &  135         &    47        \\ 
 lower surface  &   -245    & -3.2          &   82         &    33        \\ 
\noalign{\smallskip}\hline
\end{tabular}
\label{table-energy}
\end{table}

For $\alpha '-4$H we chose the points labeled
in Fig. (\ref{fig1bis}).
The most favored
adsorption site, denoted by the label BRI (in red), 
corresponds to the midpoint between two undecorated B atoms, 
while the least favored one, denoted by the 
label THB in Fig.(\ref{fig1bis}), is on top of a H atom 
forming a bond with a B atom below it.
The main results are listed in Table III.
In this structure, the minimum intersite barrier is found to be 86 K.  

If we compare the energetic data characterizing the adsorption of He
on the $\alpha '-4$H and rect-2H substrates with those found in the case
of adsorption of He on graphene,\cite{reatto_2013,moroni21} 
one can see that the binding energy in the favored adsorption site 
(HOL on graphene and rect-2H and BRI on $\alpha '-4$H) is similar 
(-298 K in He-graphene); however, both the corrugation and its anisotropy 
turn out to be more pronounced in borophane structures than in graphene, 
where the maximum corrugation (50K) and minimum intersite barrier (47K) 
are relatively small and comparable. 

\begin{table}
\caption{Binding energy, $E_b$ (from the most favored to the least favored configuration), 
for He on the $\alpha '-4$H borophane
substrate, relative to the sites shown in Fig.(\ref{fig1}).}
\begin{tabular}{|l|c|}
\hline\noalign{\smallskip}
 site       &  $E_b$(K)   \\ 
\hline
 BRI        &  -269   \\
 TBs        &  -263   \\
 TBp        &  -230   \\
 BBH        &  -224   \\
 BRIp       &  -218   \\
 TB         &  -187   \\
 HOL        &  -183   \\
 TBH        &  -174   \\
 BHB        &  -172   \\
 BH2        &  -153   \\
 THB        &  -151   \\
\noalign{\smallskip}\hline
\end{tabular}
\label{table-alpha}
\end{table}

In order to corroborate our basic results, we have repeated the 
calculations by adopting a different, vdW-corrected DFT functional, namely
the DFT-D2 functional (semiempirical in character).\cite{Grimme06}
Although with DFT-D2 the values for the energies are 
systematically more negative than those calculated with the rVV10 functional
(see Table I), so the predicted He-borophane bonding is stronger, 
nonetheless the energetic ordering is preserved, and thus we do not expect
a potential energy surface much different from that obtained using  
rVV10. In any case, from previous work,\cite{Sabatini,psil15,psil16}
the rVV10 functional turns out to be
more accurate than DFT-D2 for this kind of application.
For this reason, we decided to use the adsorption potential
generated with the rVV10 functional to perform PIMC simulations
of adsorbed $^4$He on borophane. A few additional results obtained with
the DFT-D2 potential, reported in the Supplemental Material,\cite{suppmat}
show that the main features of the adsorbed phases are preserved.
 
In addition to the lowest-energy configurations for a given investigated 
site, we have also computed the dependence upon 
the normal coordinate $z$ of the He-substrate interaction potentials.
We fitted the calculated points, for each site, using the 
results of a 8-parameter curve fitting with the form
$
\sum _{i=1}^{2} a_i \, exp(-b_i z)-\sum _{i=1}^{4}c_i/z^{2i+2},
$
whose parameters are given in Table~\ref{tab:tablepar}.

\begin{table*}[!]
\caption{
Best-fit parameters for the $z$ dependence of the He-substrate potentials
(Rect-2H structure) shown in Fig. 3.
}
\begin{center}
\label{tab:tablepar}
\begin{tabular}{lcccccccc} 
\hline
\hline    
\multicolumn{9} {c} {Rect-2H (upper surface):} \\
$ $ & $a_1$ & $b_1$ & $a_2$ & $b_2$ & $c_1$ & $c_2$ & $c_3$ & $c_4$ \\
\hline
 HOL & $-8.37576\times 10^{5}$ & $1.7212$ & $-44.0052$ & $0.0830533$ & $2.96892\times 10^{5}$ & $-9.80514\times 10^{6}$ & $3.94821\times 10^{7}$ & $-5.32952\times 10^{7}$ \\
 BRIH &  $1.57552\times 10^{4}$ & $0.79863$ & $1.54609\times 10^{4}$ & $0.798294$ & $4.73313\times 10^{5}$ & $3.30169\times 10^{5}$ & $-3.38658\times 10^{7}$ & $7.81978\times 10^{7}$ \\
 BRIB &  $-2.31914\times 10^{5}$ & $0.591921$ & $2.63864\times 10^{5}$ & $0.608549$ & $-1.40649\times 10^{5}$ & $3.03206\times 10^{7}$ & $-4.75508\times 10^{8}$ & $1.87298\times 10^{9}$ \\
 TH &  $3.86271\times 10^{4}$ & $0.371565$ & $8.47703\times 10^{5}$ & $0.920089$ & $1.12623\times 10^{7}$ & $-3.99165\times 10^{7}$ & $-3.38524\times 10^{8}$ & $6.56494\times 10^{8}$ \\
 TB &  $-5.47410\times 10^{5}$ & $1.41883$ & $-5.18545\times 10^{5}$ & $1.41895$ & $7.71972\times 10^{5}$ & $-3.95275\times 10^{7}$ & $2.26423\times 10^{8}$ & $-4.1518\times 10^{8}$ \\
 INT &  $-2.72773\times 10^{5}$ & $1.43711$ & $-1.93299\times 10^{3}$ & $0.366335$ & $-1.46259\times 10^{5}$ & $-2.92511\times 10^{6}$ & $1.23338\times 10^{7}$ & $-1.53372\times 10^{7}$ \\
\hline
\hline
\multicolumn{9} {c} {Rect-2H (lower surface):} \\
$ $ & $a_1$ & $b_1$ & $a_2$ & $b_2$ & $c_1$ & $c_2$ & $c_3$ & $c_4$ \\
\hline
 HOL & $-1.15133\times 10^{7}$ & $2.59433$ & $-8.08397\times 10^{5}$ & $1.51851$ & $5.13013\times 10^{5}$ & $-2.02464\times 10^{7}$ & $6.23524\times 10^{7}$ & $-6.64392\times 10^{7}$ \\
 BRIH &  $2.19457\times 10^{4}$ & $0.757928$ & $-5.92583\times 10^{2}$ & $0.758256$ & $4.97181\times 10^{5}$ & $-3.41012\times 10^{6}$ & $5.94483\times 10^{6}$ & $-2.08311\times 10^{6}$ \\
 BRIB &  $-1.02228\times 10^{6}$ & $1.76746$ & $-0.00970567$ & $-0.909411$ & $3.48361\times 10^{5}$ & $-1.0862\times 10^{7}$ & $4.10722\times 10^{7}$ & $-5.22449\times 10^{7}$ \\
 TH &  $-1.04288\times 10^{6}$ & $1.60283$ & $-1.34795\times 10^{7}$ & $2.80359$ & $4.83701\times 10^{5}$ & $-1.80164\times 10^{7}$ & $5.42601\times 10^{7}$ & $-5.73984\times 10^{7}$ \\
 TB &  $-0.0304066$ & $-0.767078$ & $-1.02223\times 10^{6}$ & $1.78369$ & $3.36546\times 10^{5}$ & $-1.03101\times 10^{7}$ & $3.87033\times 10^{7}$ & $-4.86229\times 10^{7}$ \\
 INT &  $-9.31143\times 10^{2}$ & $0.762755$ & $2.12973\times 10^{4}$ & $0.762542$ & $4.75234\times 10^{5}$ & $-3.43202\times 10^{6}$ & $7.85485\times 10^{6}$ & $-6.02331\times 10^{6}$ \\
\hline
\hline
\end{tabular}
\end{center}
\end{table*}

In order to perform a many-body computation one needs the full adsorption potential
$V_{He-s}(x,y,z)$. To this end we
approximate the potential $V_{He-s}$
by using a truncated Fourier
expansion over the ${\bf G}$ vectors of the two-dimensional reciprocal
lattice associated with the lattice of the substrate:

\begin{equation}
V_{He-s}({\bf r})=V_0(z)+\sum _{i=1}^{5}[V_i(z)\sum _j e^{i{\bf G}_i^{(j)}\cdot {\bf R}}]
\label{potr}
\end{equation}
where the ${\bf G}_i^{(j)}$ ($i=1,5$) are
the first five stars of two-dimensional reciprocal
lattice vectors, ${\bf G}=n_1{\bf b}_1+n_2{\bf b}_2$
($n_1,n_2=0,\pm 1,\pm 2,\pm 3,...$ and
${\bf b}_1=(2\pi/a)\hat {x}$,
${\bf b}_2=(2\pi/b)\hat {y}$ ).

The Fourier components $V_i$ 
can be easily obtained, for a given value of $z$, from
the calculated values of $V_{HOL}(z),V_{BRIH}(z),V_{BRIB}(z),V_{TH}(z),V_{TB},V_{INT}(z)$,
as defined above.\cite{nota}

\begin{figure}[h]
\includegraphics[width=8cm]{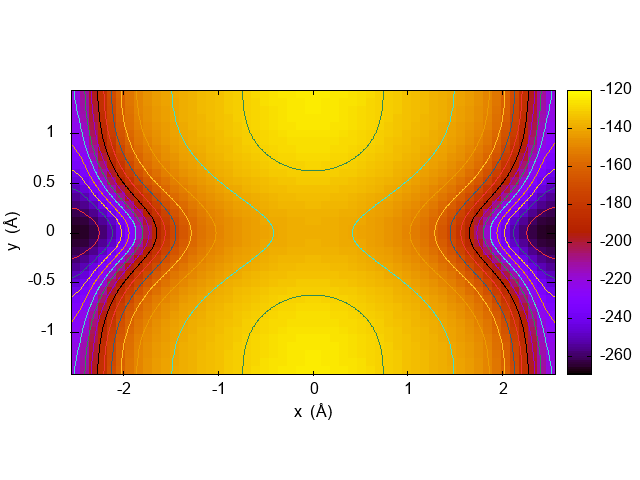}
\caption{Equal energy contours relative to the potential energy map in the unit cell 
for the {\it upper} surface of Rect-2H borophane.
}
\label{fig4}
\end{figure}

\begin{figure}[h]
\includegraphics[width=8cm]{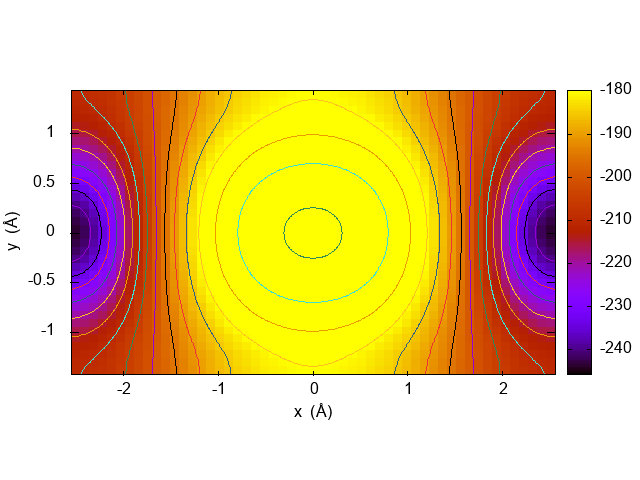}
\caption{Equal energy contours relative to the potential energy map in the unit cell 
for the {\it lower} surface of Rect-2H borophane.
}
\label{fig5}
\end{figure}

In Fig.(\ref{fig4}) and Fig.(\ref{fig5}) we show, by means of equal energy contours, the 
potential energy map in the unit cell for the upper and lower surface of the 
Rect-2H borophane monolayer, 
corresponding to the minimum value of the potential for each point $(x,y)$.
These figures clearly
show the high ridge of the adsorption potential in 
the center of the unit cell, a ridge that
divides the regions of the preferential adsorption sites, 
forming a kind of channel for easy
motion of the adsorbed atoms.

\section{Quantum Monte Carlo simulations}
\label{sec_qmc}
We use the path integral Monte Carlo method\cite{pimc} in the
worm algorithm\cite{worm} implementation to calculate unbiased thermal 
averages in the grand canonical ensemble for an assembly of ${^4}$He atoms
adsorbed on the upper surface of Rect-2H borophane (hereafter BPH-rect), 
with periodic boundary 
conditions in 
directions $(x,y)$ parallel to the substrate. 
The size of the simulation cell is $5a$=25.575\AA{} along $x$ and $9b$=25.695\AA{}  along $y$.
The interaction between He and BPH-rect is the upper-surface rVV10 potential of 
Sec. II,\cite{dft-d2} 
and the pairwise He-He interaction is the HFDHE2 Aziz potential.\cite{aziz}
In case BTP-rect could be produced as free standing sheet also the lower surface should
be relevant.
Some results for the lower surface are reported in the 
Supplementary Material.\cite{suppmat}
The technical details of the simulation are the same as 
in Ref.~\onlinecite{moroni21}.

\begin{figure}[h]
\includegraphics[width=8cm, trim = 2cm 2cm 2cm 1.5cm]{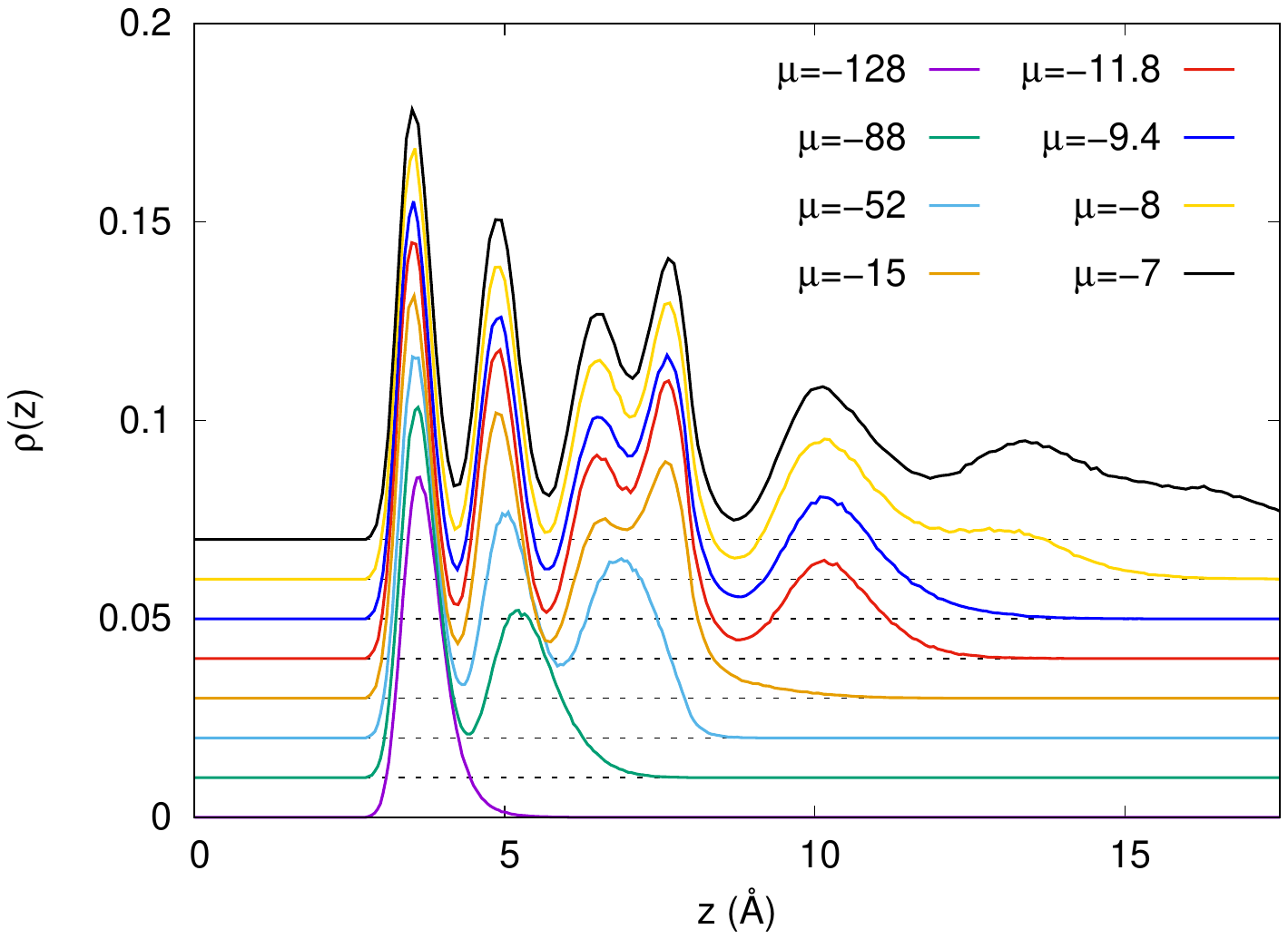}
\caption{Density of $^4$He averaged over $(x,y)$ as a function of the 
distance $z$ (in \AA) from the substrate for representative values of the chemical potential
at $T=0.5$ K. 
The integral of each curve is the average number of ${^4}$He atoms
divided by the area of the simulation cell in the $xy$ plane.
Vertical shifts are added for clarity. 
}
\label{fig:rhoz}
\end{figure}

The growth of the film proceeds by successive filling of the adsorption layers.
Individual layers are well characterized, at least at low coverage, 
by well--resolved peaks of the lateral average of the $~4$He 
density as a function of $z$, see Fig.~(\ref{fig:rhoz}).

\begin{figure}[h]
\centering
\includegraphics[width=8cm, trim = 3.5cm 2cm 2cm 3cm]{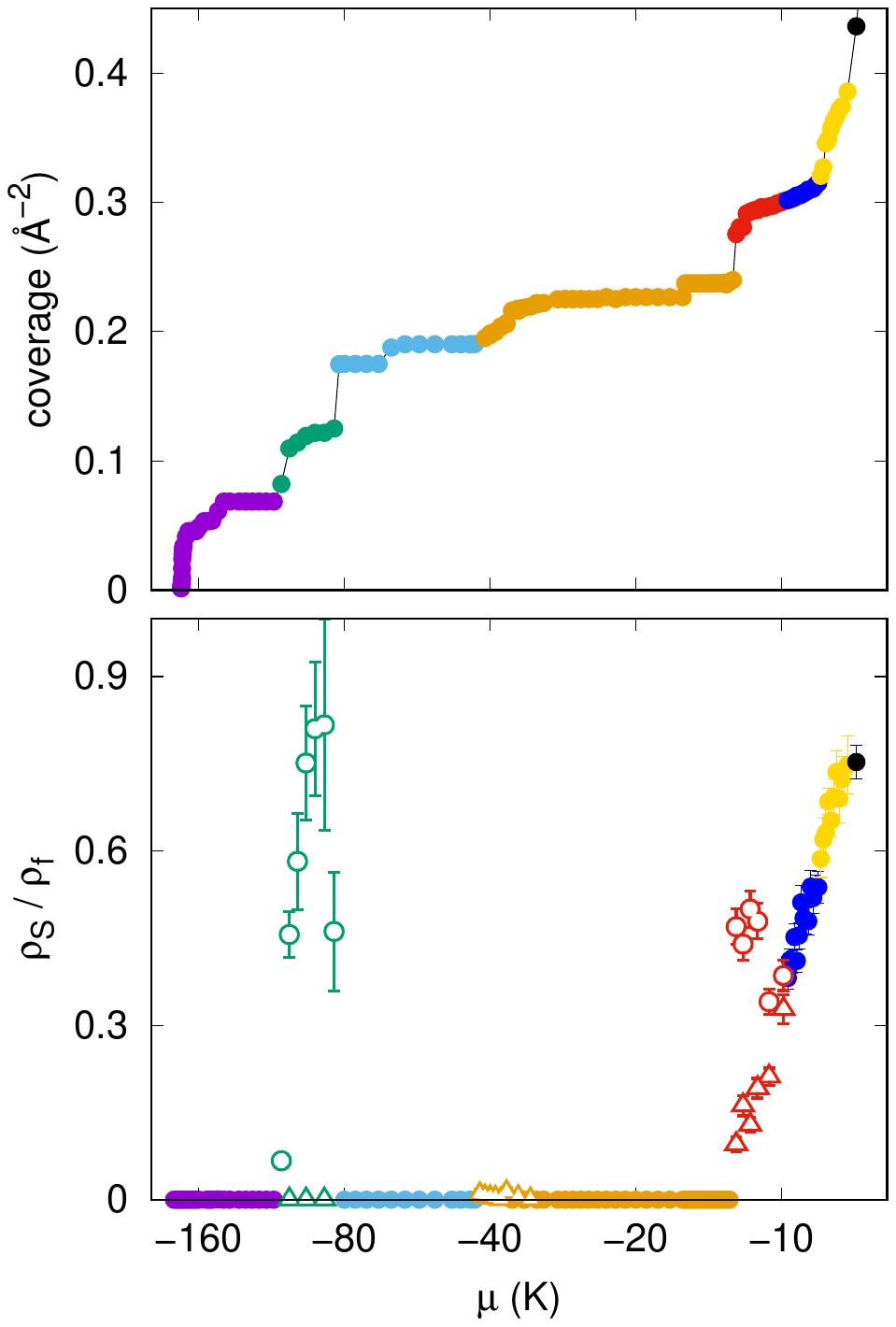}
\caption{Coverage (top panel) and superfluid fraction (bottom panel) of
${^4}$He adsorbed on the upper surface of BPH-rect as a function 
of the chemical potential at $T=0.5$ K. Different colors correspond to 
different layers. In the bottom panel, open triangles and open circles indicate
the superfluid fraction along the $x$ and $y$ directions, respectively.
Isotropic superfluid fractions are indicated by filled circles.
}
\label{fig:rho_of_mu}
\end{figure}

Figure (\ref{fig:rho_of_mu}) shows the evolution of the coverage and of the
superfluid fraction as the chemical potential $\mu$ increases, with the 
temperature kept fixed at $T=0.5$ K. Here the coverage is defined through 
the number $N$ of ${^4}$He atoms in the simulation cell.
The main jumps or shoulders in $N(\mu)$, marked by changes in color,
indicate promotion to the next layer (except red to blue; see below). 
Flat portions of the $N(\mu)$ curve correspond to crystalline ordering. 
The superfluid density $\rho_s$ is calculated with the winding number
estimator.\cite{pimc}
The superfluid fraction shown in Fig. (\ref{fig:rho_of_mu}) is defined as $\rho_s/\rho_f$, where
$\rho_f=\rho-\rho_{cryst}$ is the total density $\rho$ of $^4$He minus the
density $\rho_{cryst}$ of the crystalline layers not involved in the
superfluid flow. In detail, we subtract $N_{cryst}=45$ atoms for layer 2
and $N_{cryst}=159$ for layer 5 and above.

When the superfluid fraction is anisotropic in the
$(x,y)$ plane, we show separately the contributions coming from the $x$ 
and $y$ components of the winding number.\cite{pimc}

\begin{figure}[h]
\includegraphics[width=8.5cm, trim = 3cm 5cm 3cm 2cm]{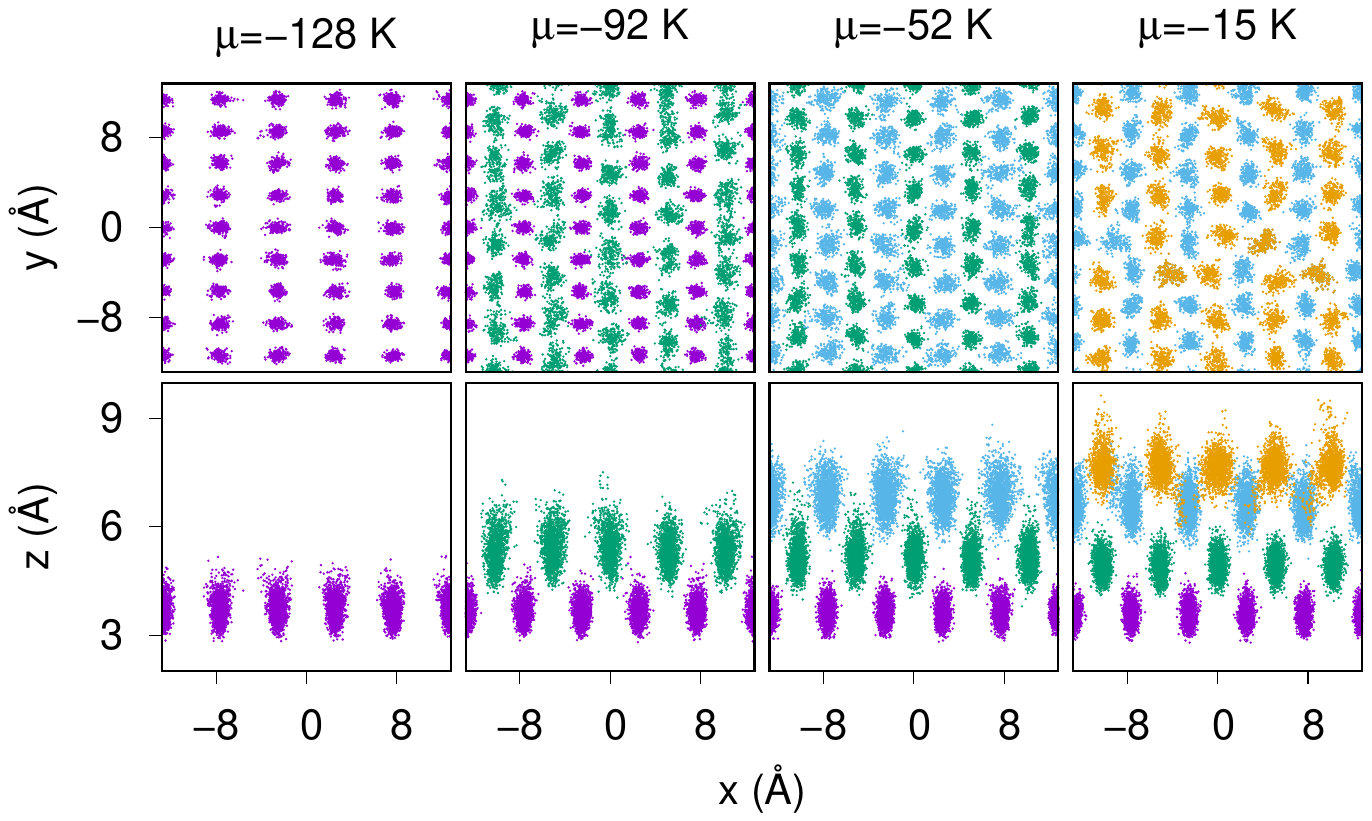}
\caption{Snapshots of ${^4}$He atoms adsorbed on BPH-rect
for various coverages.
Each particle is represented by a sequence of ``beads''\cite{pimc} (shown here 
with a stride of five). Upper panels are top views, Lower panels are
$(x,z)$ side views. Different colors correspond to different layers.
Individual atoms are assigned to a particular layer based on the
vertical position of the centroid of its path.
Layers behind the second topmost are not shown in the top views.
Note the expanded scale for the $z$ component in the side views.
}
\label{fig:snapshots_1}
\end{figure}

For representative values of the coverage, Figs.~(\ref{fig:snapshots_1})
and (\ref{fig:snapshots_2})
show snapshots of the path integral configurations, which 
encode information on the localization (or lack thereof) of the
$^4$He atoms. 
The deep valleys in the adsorption potential
along the $y$ direction above the HOL sites seen in Fig.~(\ref{fig4}), 
combined with their relatively wide separation of $a=5.115$\AA ~along $x$, 
deeply influence the structure of the first several layers:
The atoms in the first layer are tightly confined above the HOL sites,
and because of the He-He hard-core repulsion they
constitute high ridges for the subsequent adatoms; therefore the second 
layer atoms are confined between lines of first-layer atoms, and they 
constitute in turn high ridges for the next layer. This interlocking structure 
softens with increasing coverage, but it persists up to the fifth layer. 
This is completely different from, say, the prototypical case of graphite
\cite{manousakis}, where the effect of corrugation is already very
small for the second layer.\cite{boninsegni}

\begin{figure}[h]
\includegraphics[width=8.5cm, trim = 3cm 5cm 3cm 2cm]{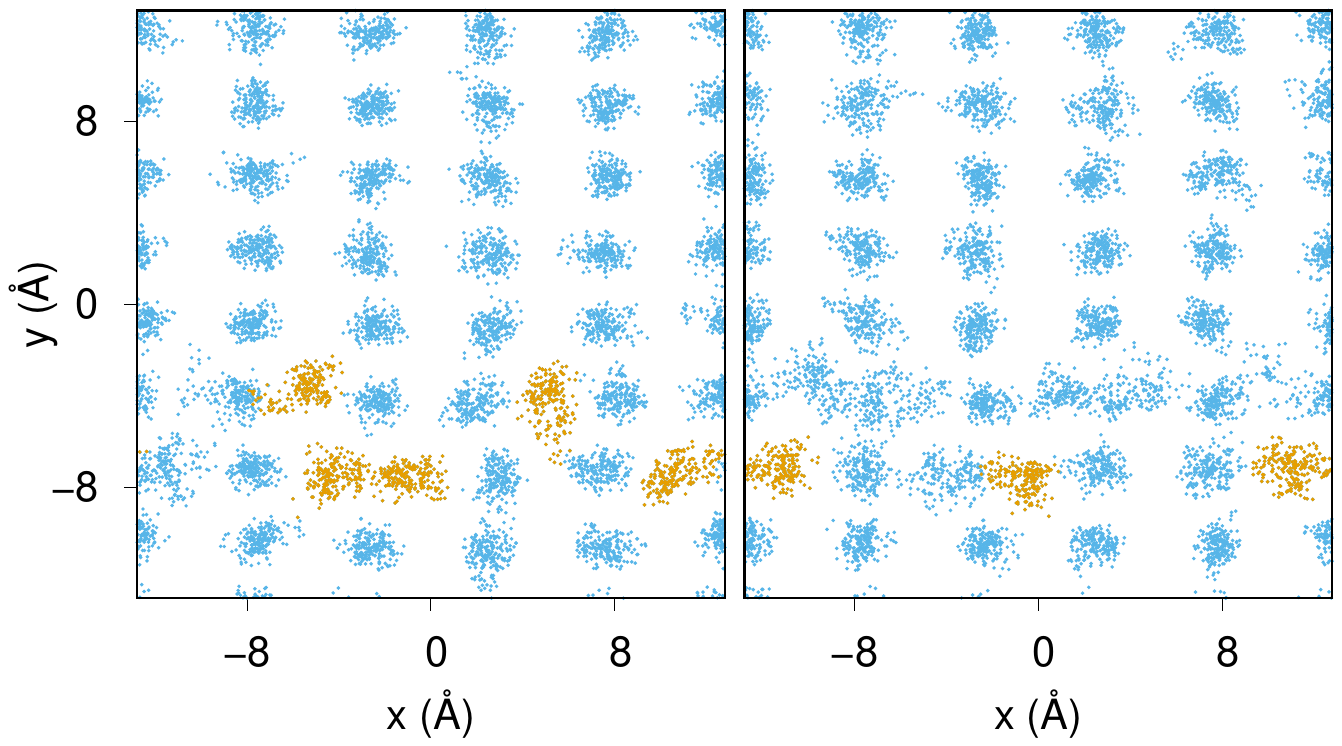}
\caption{
Snapshots of the third- and fourth-layer ${^4}$He atoms (light blue and
orange, respectively) 
adsorbed on BPH-rect for $\mu=-40$~K.
The configuration shown in the right panel
includes a path with
non-zero winding number around the
$x$ direction involving third-layer atoms.
}
\label{fig:filamento}
\end{figure}
\begin{figure}[h]
\includegraphics[width=8.5cm, trim = 3cm 5cm 3cm 2cm]{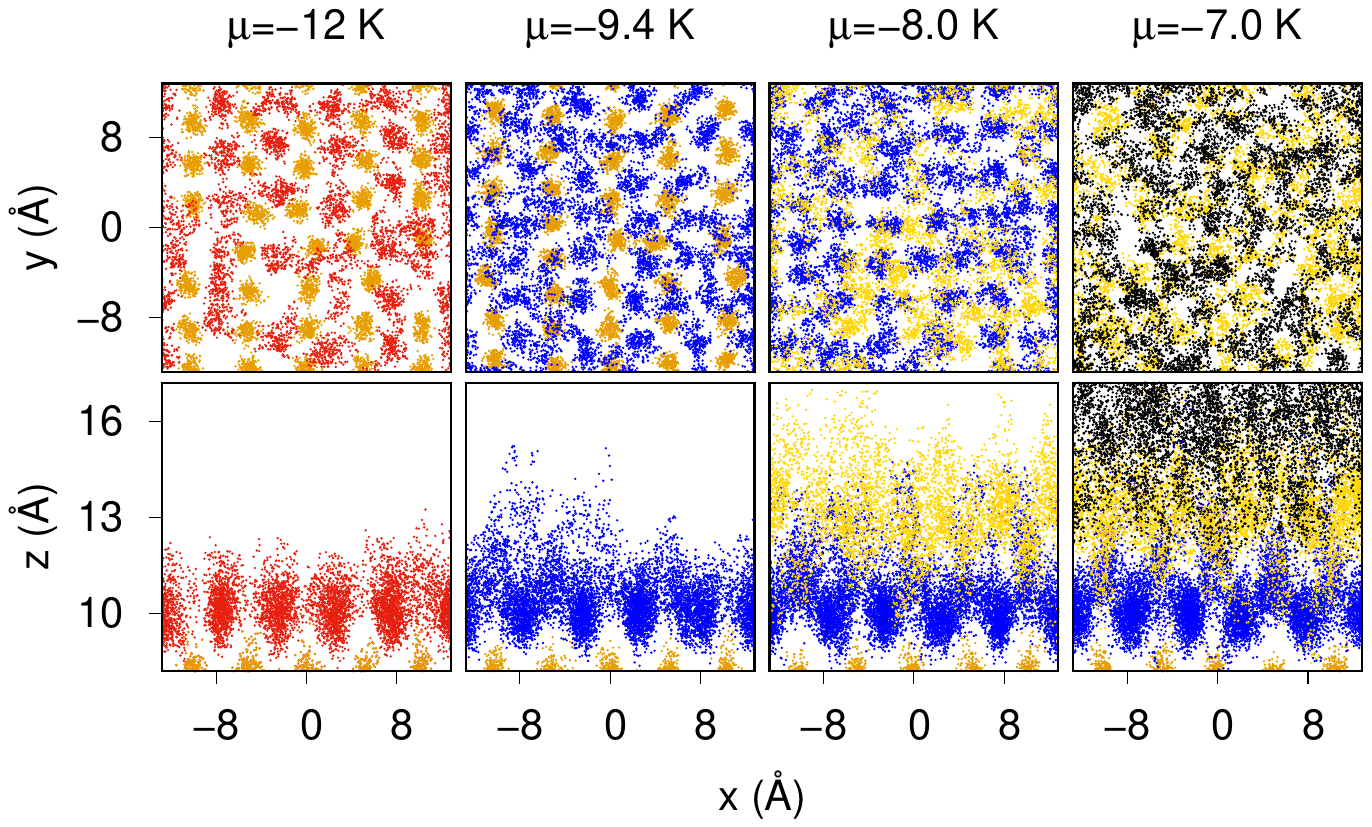}
\caption{Same as Fig. (\ref{fig:snapshots_1}) for higher coverages.
Deviating from the code of one color per layer, red and blue atoms 
correspond to different coverages within the fifth adsorption layer.
The assignment of (red)blue color is based on the (an)isotropy of the
superfluid flow rather than on the distance of the centroids from
the substrate, see text.
}
\label{fig:snapshots_2}
\end{figure}
\begin{figure}[h]
\includegraphics[width=8cm, trim = 2cm 2cm 2cm 2cm]{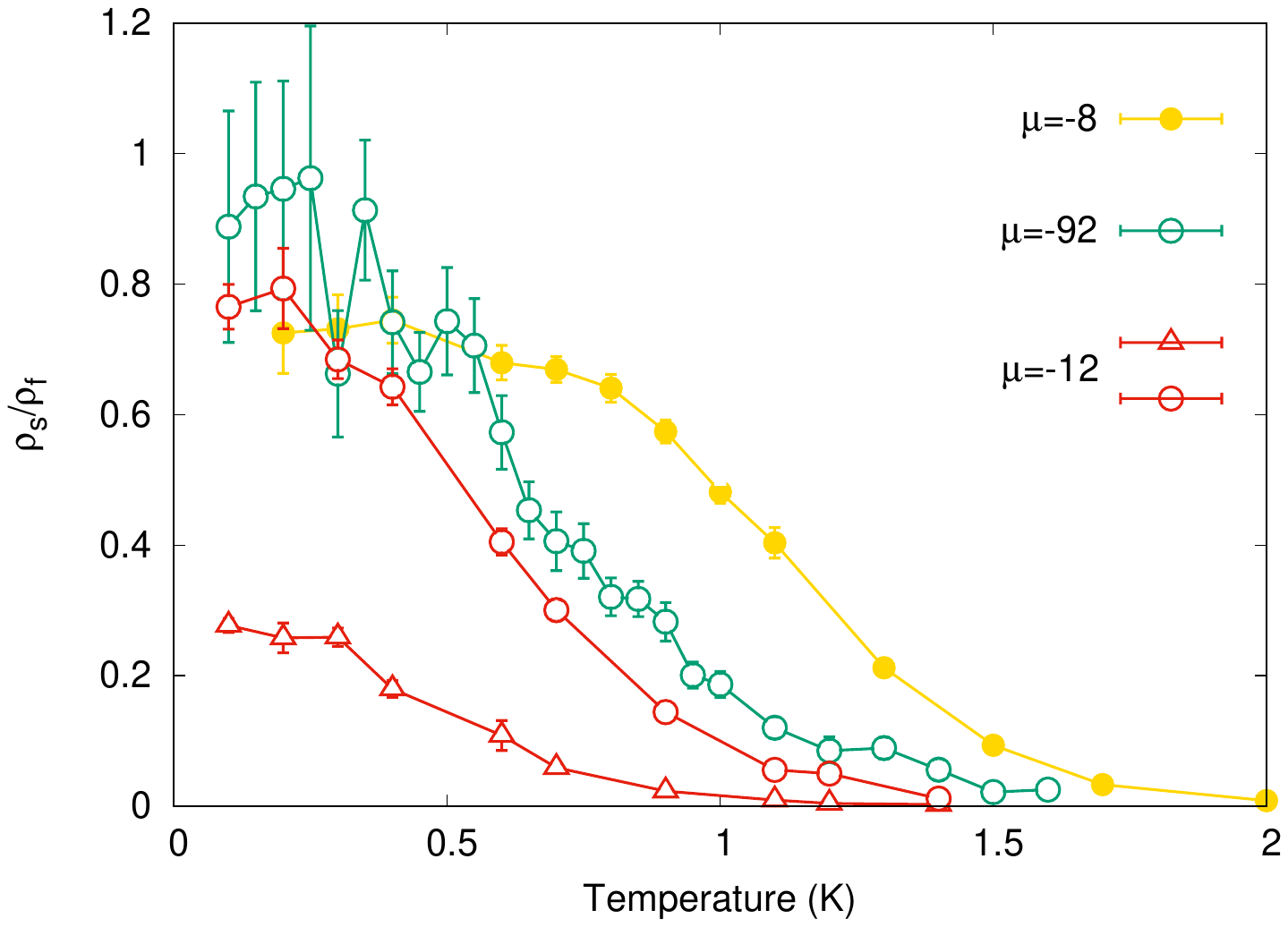}
\caption{Dependence on the temperature of the superfluid fraction
for selected values of the coverage. 
Open triangles and open circles indicate
the superfluid fraction along the $x$ and $y$ directions, respectively.
Isotropic superfluid fractions are indicated by filled circles.
}
\label{fig:rhos_of_t}
\end{figure}

We now discuss the properties of individual adsorption layers.\\
{\em Layer 1.} 
For coverages up to $N=45$ atoms per simulation cell, each ${^4}$He atom is 
pinned to one 
of the HOL absorption sites; see the leftmost panels in Fig.~(\ref{fig:snapshots_1}).
At complete
filling of the first layer, we have an ordered state in registry with the substrate. 
It can be seen
that the distance 2.855\AA ~between neighboring $^4$He atoms along the $y$ direction
corresponds to the interatomic distance in solid $^4$He at a very high pressure. 
Here, this large
compression is due to the very strong adsorption potential.
When 
filling is incomplete, particles can hop between sites in the $y$
direction and, in principle, sustain a superfluid flow. However, the intersite
minimum density along $y$ is as small as 1/150 the peak value, so that a very 
low critical temperature is expected.
Indeed, we find no superfluidity down to the lowest temperature studied, 
$T$=0.1~K.
For incomplete fillings, the vacant adsorption sites tend to align along the $x$
direction, in order to minimize the energy contribution from the attractive tail of the
interaction between atoms $\sim a$ apart.\cite{suppmat}

{\em Layer 2.} The second layer is a fluid distributed in separated channels 
along the $y$
direction, alternate to the HOL sites, as shown by the panels of Fig.~(\ref{fig:snapshots_1})
corresponding to $\mu=-92$~K. Note in particular that the path of nearest neighbor particles
are rather delocalized along $y$ and often overlap. Further support to the assignment
to a fluid phase comes from the static structure factor $S({\bf k})$, which 
features
characteristic ridges.\cite{moroni21,suppmat} Each channel contains up to 7 $^4$He atoms, i.e.
the interatomic distance along $y$ is dictated by the He-He repulsion rather than by the 
9-fold periodicity of both the substrate corrugation and the first layer density profile. 
This mismatch is instrumental in frustrating a possible solid order and stabilizing instead
the observed fluid phase.
The superfluid fraction is non zero, in the finite samples simulated. It is entirely due 
to intra-channel exchanges, because the $x$ component of the winding number is zero. 
It can be
noticed that the superfluid fraction in this second layer has a maximum as function of
chemical potential and $\rho_s$ drops as $\mu$ approaches the value for promotion to the next
layer. This drop can be considered as a precursor of the crystallization of the particles of
the second layer when the third layer is formed.
A similar situation has been studied through PIMC simulations of strongly interacting 
2D bosons confined to 1D channels by an external periodic potential modulated along one 
direction.\cite{yao} Below quantum degeneracy, a 2D-1D dimensional crossover is observed 
as the amplitude of the external potential increases at fixed particle density. 
The coupling between 
1D channels is further characterized as either coherent or 
incoherent\cite{yao} when the transverse superfluid fraction has a finite or vanishing 
value, respectively. Based on this criterion, the $^4$He atoms in the second layer 
belong to the 1D incoherent regime.

{\em Layer 3.} The promotion to the third layer is very sharp. For sufficiently low
temperatures\cite{suppmat} we observe the sudden appearance
of a 7$\times$5 lattice with $x$ coordinates alternate to the second layer channels,
i.e. back on top of the HOL sites
(panels at $\mu=-52$~K in Fig.~(\ref{fig:snapshots_1})). Concurrently, this lattice pushes 
the density of the second layer closer to the substrate, as shown by the shift of 
the second peak in Fig.~(\ref{fig:rhoz}), and strongly localizes the paths of the
second layer in the $y$ direction; thus the second layer crystallizes as well and
the superfluid density drops to zero. 
The pale blue portion in the upper panel of Fig. 8
is flat, except for a jump of the coverage between 0.175 and 0.190\AA$^{-2}$,
corresponding to N=115 and N=125 $^4$He atoms in the simulation cell;
such a jump corresponds to
a switch between the 7$\times$5 lattice to an $8\times$5 arrangement, both in the third
and in the second layer. In the thermodynamic limit, the second and third layers 
presumably evolve as matched compressible crystals, incommensurate to the
corrugation of the substrate in the $y$ direction.  
In principle, this could be verified by an experimental measurement of $S({\bf k})$.

{\em Layer 4.}
The alternation in the $x$ direction with the atoms of the previous layer
continues in the fourth layer, but it is much less sharp, as it is the 
separation among the paths of individual atoms, see 
the bottom right panel of Fig.~(\ref{fig:snapshots_1}).
The low coverage regime of this layer, for $\mu$ between $-41$ and $-37$~K, 
features a very peculiar structure:
the promoted atoms disrupt the underlying third-layer crystal, forming 
one disordered stripe that stretches across the simulation 
cell and support a small superfluid signal ($\rho_s/\rho_f\lesssim 0.01$)
along the $x$ direction, as shown for
two configurations at $\mu=-40$~K in Fig.~(\ref{fig:filamento}).
For higher coverages the stripe disappears as the atoms come to be
evenly distributed along the $y$ direction.
Despite not being fully ordered, 
the system presents neither liquid-like ridges in the 
structure factor, nor superfluid flow. We therefore assign this
higher-coverage regime of the fourth layer
to crystal phases.
The density is lower than in the third layer at completion: 
in the example shown in Fig.~(\ref{fig:snapshots_1}) at $\mu=-15$~K, there 
are 34 atoms in the fourth layer.
At this stage, layers 3 and 4 are best seen as
a single buckled crystal, not
commensurate to the substrate in the $y$ direction.
As the coverage increases, this crystal
evolves through two different structures, not straightforwardly identified
due to the presence of several defects.
In the Supplemental Material\cite{suppmat} we show that
such defects are due to the geometry of the simulation cell, 
unable to accommodate the preferred crystalline structure(s),
and we address the underlying lattices.
Also in this case, a potential measurement of $S({\bf k})$ could
be directly compared with the Bragg peaks of the predicted structures.

{\em Layer 5.} In discussing the fifth layer we deviate from the
code of one color per layer. The initial coverage of layer 5 is illustrated
in red in Figs.~(\ref{fig:rhoz},\ref{fig:rho_of_mu},\ref{fig:snapshots_2}).
It is a liquid modulated along the $x$ direction, although more weakly
than all the structures previously described, and very delocalized in the
$y$ direction. The superfluid fraction in this layer is non-zero not only
in the $x$ direction but also in the $y$ direction and in this initial coverage of layer 5 $\rho_s$
is strongly anisotropic with a ratio of the two components that is as large as three.
When the coverage increases (blue color, $\mu\ge -10.2$~K), 
the fluid channels along the $y$ directions overlap significantly,
and the superfluid fraction becomes isotropic despite the persistence
of a residual modulation along $x$, particularly closer to the substrate.
From the point of view of superfluidity, 
layer 5 offers
an interesting example of crossover between two regimes:
at small coverage of this layer the superfluidity is strongly
anisotropic reflecting the ridges of the adsorption potential 
of the preplated substrate
and this regime merges into an isotropic one in the
$(x,y)$ plane as the coverage of this layer moves toward
completion.

{\em Layers 6 and 7.} 
For higher coverages the modulation along $x$ fades away, and
the distinction between layers gets blurred. Promotion to the sixth layer onwards
does not induce crystallization of layers beneath, and the film evolves smoothly towards
the bulk system.

The dependence of $\rho_s/\rho_f$ on 
temperature is illustrated in 
Fig.~(\ref{fig:rhos_of_t}).
We have not attempted to determine the critical temperature by finite size scaling.
Qualitatively, we see that superfluidity persists at higher temperatures for a nearly 
homogeneous density ($\mu=-8$~K) than for strongly modulated ones ($\mu=-12$ and $-92$~K).
The superfluid fraction tends to flatten as a function of 
temperature below 0.2-0.4 K depending on the coverage, so that
the results at the lowest $T$ of the present computations
can be considered 
representative of the ground state
of the system. In a uniform superfluid the superfluid
fraction should be unity at $T=0$~K but this is not the
case for a non-uniform superfluid as shown by Leggett\cite{leggett}
and the larger is the density modulation the stronger is
the depression of the superfluid fraction. This 
reduction
has been experimentally verified with cold bosons atoms
in an optical lattice\cite{tao} and it has been theoretically predicted 
in the case of $^4$He adsorbed on fluorographene\cite{moroni21} and
in a Bose-Einstein condensate in an optical lattice.\cite{ar} 
In the present study of $^4$He on BPH-rect we find a 
reduction of the superfluid fraction in the adsorbed layers that
show superfluidity, i.e. in the second layer, in the fifth
and above even at the lowest $T$ of our computations.
The strongest reduction is found in the superfluid 
fraction in the $x$ direction, i.e. across channels, in the fifth
layer at coverage with anisotropic superfluidity. In fact,
this is the case with the largest density modulation, as
shown in the Supplemental Material.\cite{suppmat} We use the ratio
$f=\rho_{min}/\rho_{max}$ of the minimum to the maximum density
as a measure of the amplitude of the density modulation.
In the $y$ direction $f$ is 0.47 in the first layer at $\mu=-92$~K
and it is essentially zero in the $x$ direction. In the fifth
layer at $\mu=-12$~K we find $f=0.45$ in the $y$ direction along
the channels and $f=0.11$ in the $x$ direction, indeed the
case with the strongest reduction of the superfluid 
fraction. Also in the case of isotropic superfluidity we find
some reduction of the superfluid fraction from unity as
can be seen in Fig. (\ref{fig:rhos_of_t}) for $\mu=-8$~K. This presumably is
due to a remnant density modulation in the layers rather
close to the substrate.

\section{Summary and conclusions}
In this paper we have investigated whether borophane is an intersting
substrate for the study of adsorbed $^4$He in the search 
of superfluid phases in restricted geometry.
We have characterized the adsorption properties of $^4$He
atoms on two allotropes of borophane, $\alpha '-4$H and
Rect-2H, by using ab initio DFT. 
Compared to graphene/graphite,
the adsorption energy of a He atom on both allotropes of borophane is similar,
but the corrugation of the adsorption potential is much stronger
and it has a more structured shape reflecting the presence and positions 
of the chemisorbed hydrogens. 

In the case of $\alpha '-4$H borophane the 
lowest energy barrier for moving from one adsorption site to a neighboring 
one is about 80\% larger than in the case of graphene/graphite.
The maximum corrugation of the adsorption 
potential, defined as the difference between its minimum and
its maximum value in the first adsorption layer, is more than 100
K, to be compared to about 3 K in the case of graphene.
On the basis of such features of
the adsorption potential and comparison with the case
of graphene/graphite we expect that the $^4$He atoms will
be localized around the adsorption sites. 
Such adsorption sites on $\alpha '-4$H borophane form a triangular 
lattice so that at low temperature the preferred state for the first adsorbed
layer should be a registered triangular state similar to the
well known $\sqrt{3}\times\sqrt{3}{\rm R}30^{\circ}$ phase on 
graphite/graphene. 
Despite this similarity, there are important symmetry differences:
the adsorption potential around an adsorption site is invariant for 
rotations of $60^\circ$ for graphite/graphene,
whereas in the case of $\alpha '-4$H borophane 
it has no symmetry for rotation,
as can be seen from Fig. 4.
Therefore, also considering the stronger amplitude of the corrugation,
we expect that a study of $^4$He adsorbed on $\alpha '-4$H at 
higher coverages (not pursued here) could unveil novel features.

The adsorption potential of He on Rect-2H borophane
is characterized by parallel deep channels with large 
energy barriers (of the order of 135 K) between neighboring
channels and much smaller barriers (of the order of 47 K)
along a channel. As a consequence, the first few 
layers of adsorbed $^4$He have a pronounced one--dimensional
character. Our results show that this substrate offers a
unique platform to study degenerate bosons and 
superfluidity in different dimensions.

In the first adsorbed layer quantum degeneracy is
present in only one direction, along a channel, but 
presumably superfluidity might be present only at 
temperatures well below those of our simulations. In the second
layer quantum degeneracy shows up as a one-dimensional
superfluid response on the length scale of our simulated
system; this represents a 1-D incoherent regime because
particles in neighboring channels are not phase correlated.
The first few atoms of the fourth layer disrupt the underlying
crystalline structure, combining with third-layer atoms to form
a compelling disordered stripe along the $x$ direction.
In the fifth layer we find two regimes. At low 
coverage the system is a strongly anisotropic 2-D superfluid,
the superfluid fraction in the direction perpendicular to
the channels is about one third of the fraction along the
channels. This anisotropy diminishes as the coverage of
the layer increases and at a critical value the 2-D 
superfluidity becomes isotropic in the $(x,y)$ plane.
Starting from the sixth
layer, the system evolves towards a bulk 3-D superfluid.

The present
study suggests a number of possible extensions. For 
instance, in the case of Rect-2H borophane a systematic
size scaling study will give information on criticality in
isotropic versus anisotropic superfluidity in 2D in the fifth
adsorbed layer or, in the case of 1-D incoherent 
superfluidity of the second layer, it will allow a comparison
with the Luttinger fluid paradigm,\cite{haldane,giamarchi} a 1-D model from
which the present system deviates because the bosons
move in a periodic potential along a channel and because
rows of atoms in neighboring channels, even if phase 
uncorrelated, are spatially correlated. Also, a study of $^4$He
on $\alpha '-4$H borophane at coverage beyond the here 
predicted registered triangular phase should be of interest.
The study of the fermion $^3$He adsorbed on graphite has
been a fruitful field of research.\cite{greywall_1990_1991} 
The present results for
the interrelation between quantum degeneracy and 
dimensionality in the case the boson $^4$He on Rect-2H 
borophane suggest the interest of a similar study for $^3$He.

\noindent
\narrowtext

\newpage

\section{Supplemental Material}
This Supplemental Material contains\\ 
(i) examples of vacancy alignment in the first adsorption layer;\\
(ii) the static structure factor with two adsorption layers, with a fluid component,
and three layers, entirely solid;\\
(iii) the formation of the solid third layer;\\
(iv) further analysis of the buckled crystals in the third and fourth layers;\\
(v) density profiles in $x$ and $y$ for the second and the fifth
adsorption layer;\\
(vi) results for the first two adsorption layers on the lower surface;\\
(vii) results for the first two adsorption layers with the DFT-D2 potential.\\
Energies are in K and distances in \AA.\\

\subsection{Vacancy alignment}

\begin{figure}[h!]
\includegraphics[width=10cm]{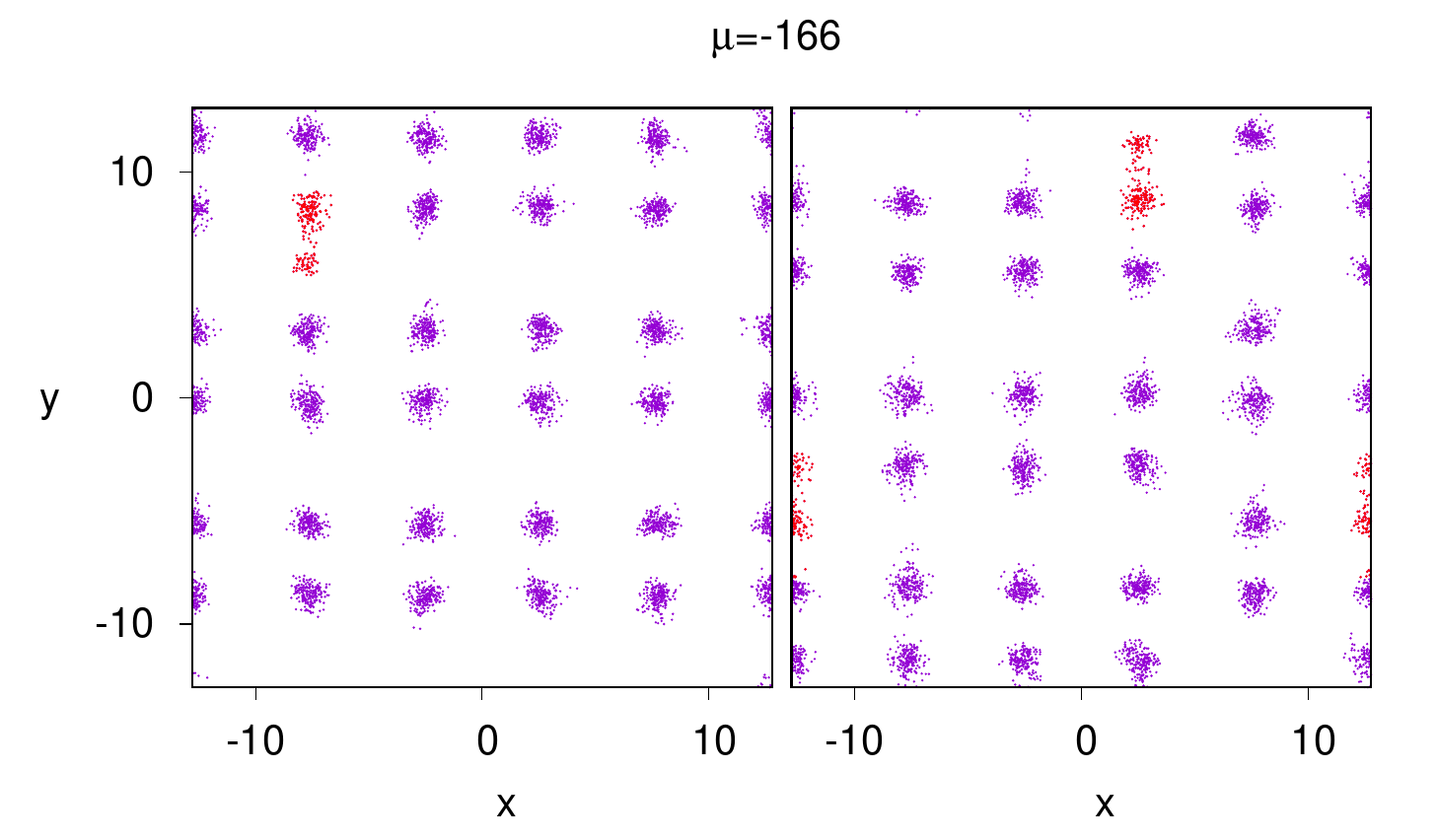}
\caption{Two snapshots of 35 $^4$He atoms partially filling the
45 adsorption sites present in the simulation cell. Each particle is
represented by a sequence of beads, shown here with a stride 
of five. Red beads denote particles hopping between two sites.}
\label{centroids_suppmat}
\end{figure}

In the first layer atoms are pinned one-on-one to adsorption sites. For
incomplete filling, vacant sites tend to align along the $y$ direction,
as shown by two examples in Fig.~\ref{centroids_suppmat}.

\subsection{Static structure factor}

\begin{figure}[h!]
\includegraphics[width=\columnwidth]{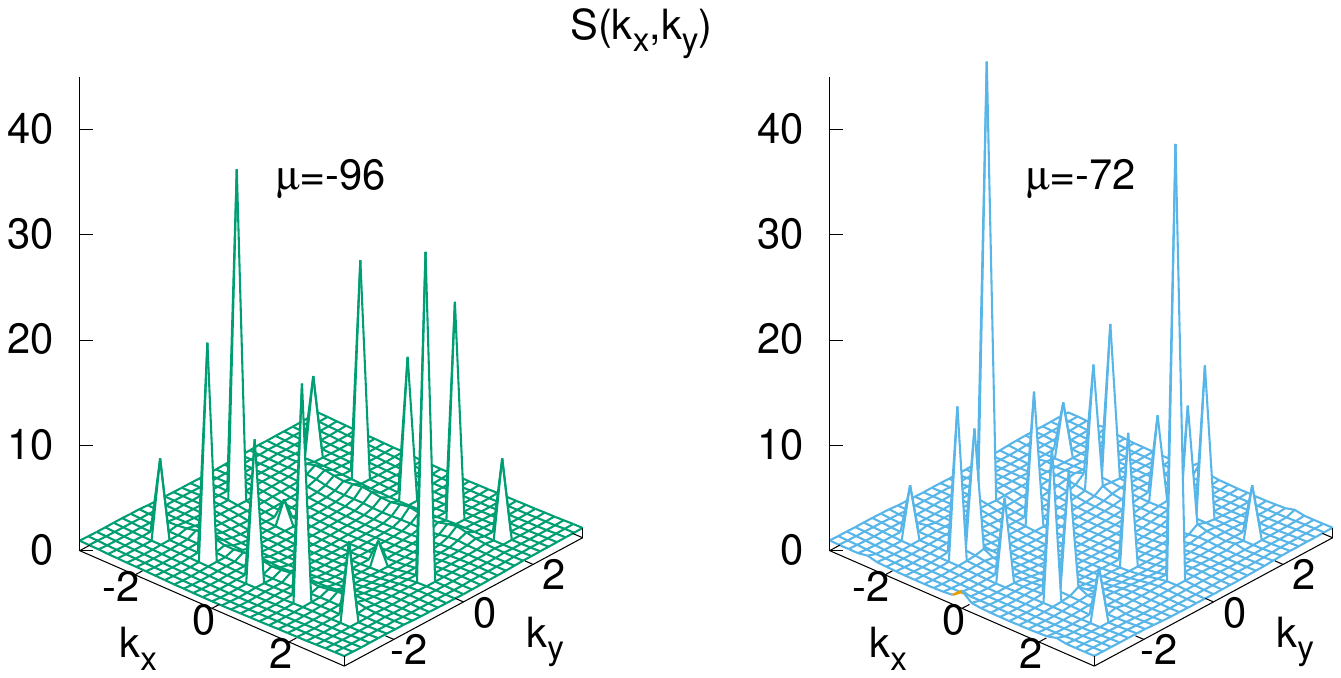}
\caption{Two-dimensional structure factor for two (left) 
and three (right) adsorption layers.}
\label{sofk_72_96}
\end{figure}

The assignment to a solid or fluid phase in a given layer is
supported by the structure factor (see Fig.~\ref{sofk_72_96}).
For two layers, alongside various Bragg peaks corresponding to the period 
$b=L_y/9$ of the substrate, there is a ridge at $k_y=\pm 1.71$ signalling
fluid behavior along the $y$ direction; in the fully solid structure 
which develops after promotion to the third layer, 
such a ridge gives way to additional Bragg peaks corresponding
to a period $L_y/7$ dictated by the He-He hard core.
 
\subsection{Formation of the solid third layer}
We show in Fig.~\ref{np_III} the coverage as a function of the
chemical potential around the density
promotion to the third layer for $T=0.5$ and $T=0.25$. 
The depth and spacing of the minima of the potential
created by the substrate and the first two layers accommodate
a commensurate crystal of third-layer atoms. The filling
of the third layer is abrupt for $T\lesssim 0.5$. For
incomplete filling at $T=0.5$ we observe coexistence between
solid and coalesced vacancies, rather than a homogeneous fluid
of vacancies (see inset of Fig.~\ref{np_III}).
\begin{center}
\includegraphics[width=\columnwidth]{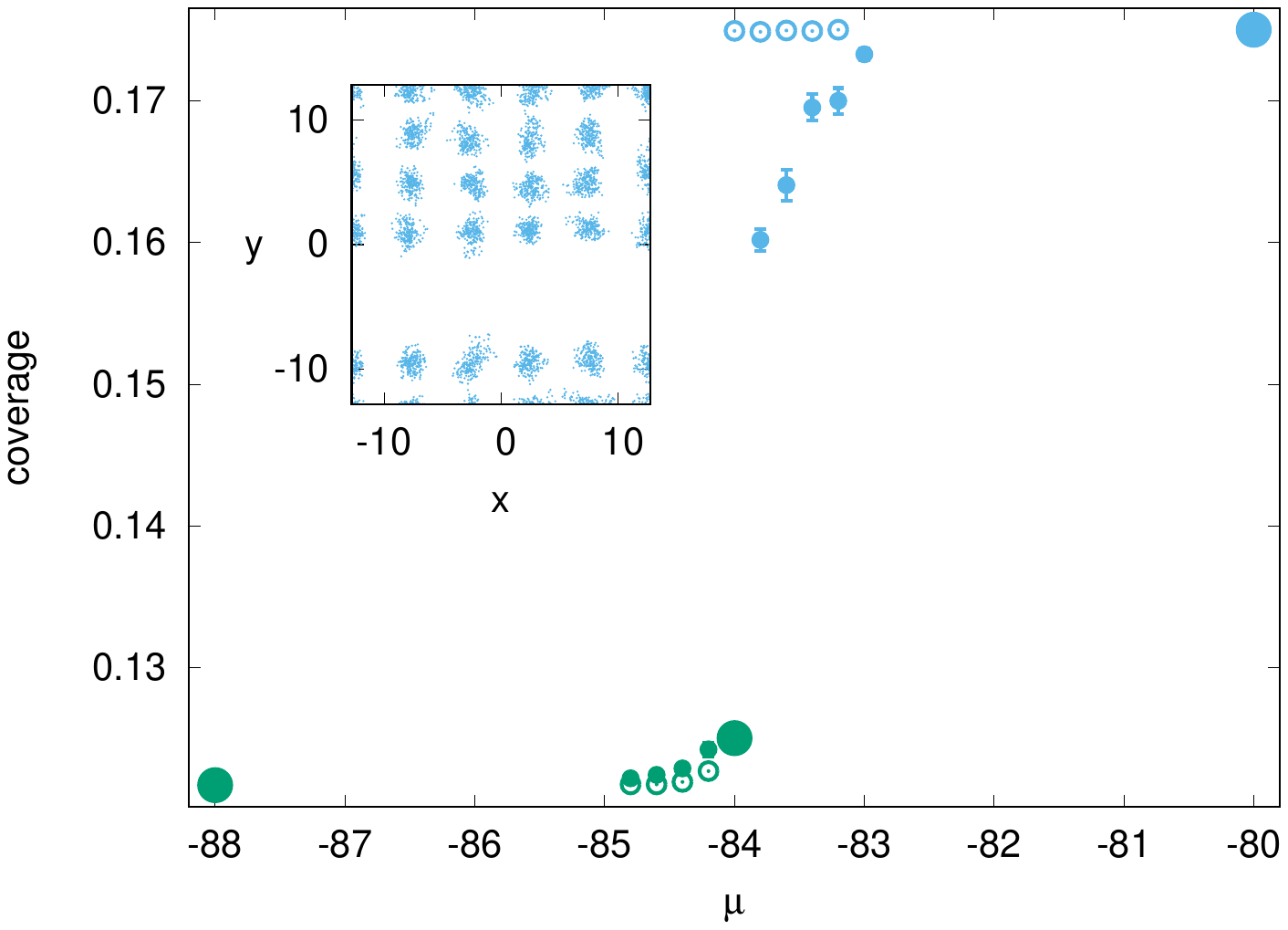}
\begin{figure}[h!]
\caption{
Coverage of $^4$He adsorbed on the upper surface of BPH-rect
on a fine grid of chemical potential around third-layer promotion
at $T=0.5$ (filled circles) and $T=0.25$ (open circles). The
larger symbols are data from Fig.~8 of the main text.
Error bars smaller than the symbol size are not shown.
The inset shows a snapshot of the paths of third-layer
atoms at $\mu=-83,5$, $T=0.5$.}
\label{np_III}
\end{figure}
\end{center}

\subsection{Density profiles in $x$ and $y$ for the second and the fifth
layer}
\begin{figure}[h!]
\includegraphics[width=\columnwidth]{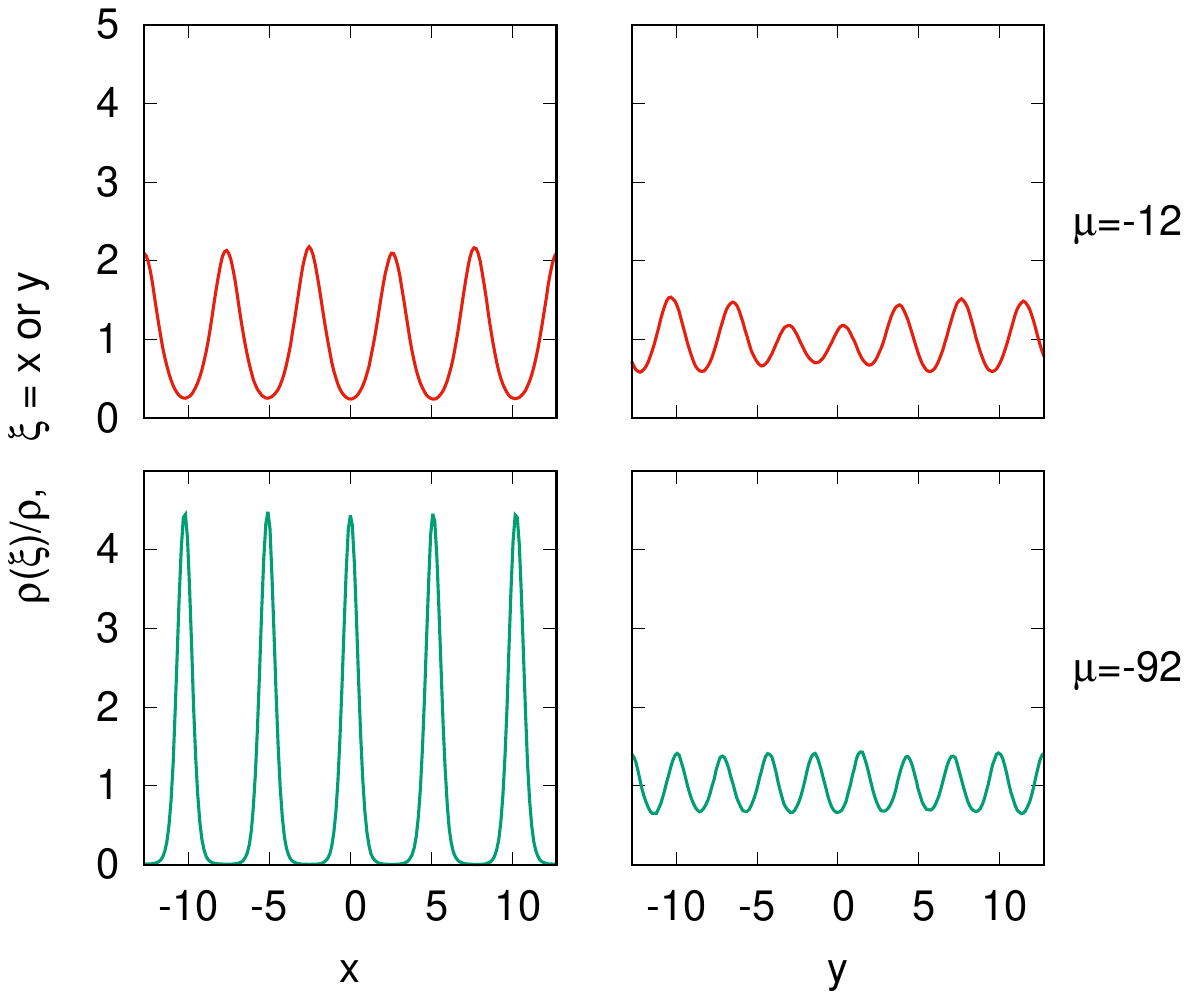}
\caption{
Density modulations of the second ($\mu=-92$, green lines) 
and the fifth ($\mu=-12$, red lines) adsorption layer.
}
\label{fig_rhoxy}
\end{figure}

In Fig.~\ref{fig_rhoxy} we show the density modulations $\rho(\xi)/\rho$
in the $x$ and $y$ directions of the second layer at $\mu=-92$
and the fifth layer at $\mu=-12$.
Here $\xi$ stands for either $x$ or $y$, $\rho$ is the average density,
and $\rho(\xi)$ is the density averaged over the two directions other than 
$\xi$. 
The non-equivalence of the minima and maxima in the fifth layer (top panels),
particularly evident in the $y$ direction,
is due to disorder present in the (localized) $^4$He atoms
in the underlying fourth layer.

\subsection{Buckled crystals in the third and fourth layers}
Starting from a sufficiently high coverage in the fourth layer,
the He atoms of the third and fourth layers combine to form a
buckled crystal, that in turn switches from a lower density structure
to a higher one upon increasing the chemical potential. 
However the results presented in the main text
do not allow us to clearly identify the lattice symmetries, due to
the presence of several defects. We are confident that such defects
are not an artifact of incomplete equilibration in the simulation: 
visual inspection of the running averages of all relevant quantities
is indicative of well-converged runs. This is strongly supported by 
the smoothness of all averages as a function of $\mu$ within the
estimated statistical uncertainties, as seen for instance in 
Fig.~(8) of the main text (note that each value of $\mu$ is addressed
through an independent simulation starting from an empty cell). 
\begin{figure}[h!]
\includegraphics[width=\columnwidth]{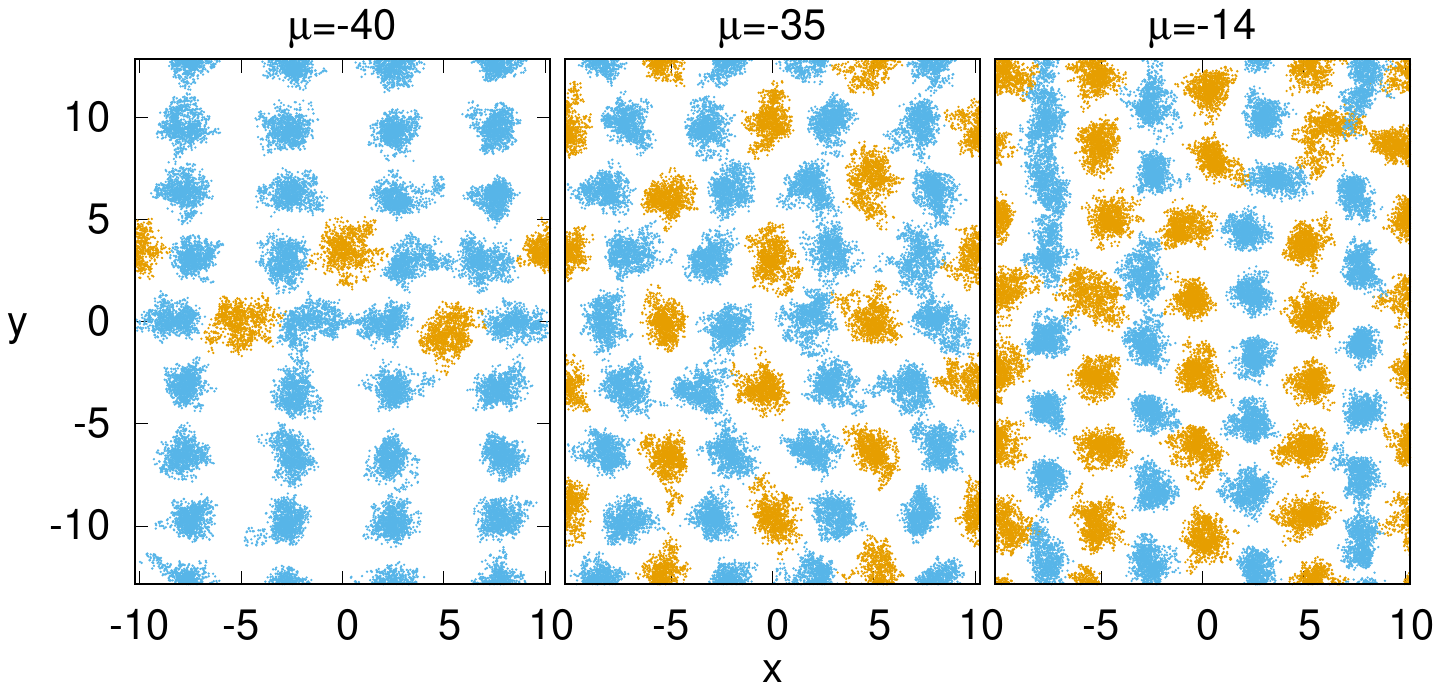}
\caption{Snapshots of $^4$He atoms in the third and 
fourth adsorption layer (light blue and orange, respectively).
}
\label{esagoni}
\end{figure}

The presence of defects is a finite size effect due to the geometry 
of the simulation cell, unable to accommodate the preferred crystalline 
structure(s). This is demonstrated in Fig.~(\ref{esagoni}) using a cell
of size $4a\times 9b$ that has an even number of adsorption sites 
in the $x$ direction.
The snapshot of the paths in the middle panel at $\mu=-35$ pertains 
to the lower-density buclked solid: there are 32 atoms in the third layer 
in an exagonal arrangement, slightly stretched in the $y$ direction, 
and 16 atoms in the fourth layer at the centers of the exagons.
The higher density solid at $\mu=-14$ features rows of atoms
along the $y$ direction, alternate between layer 3 and 4, as seen
in the snapshot of the right panel.
There are still defects: there are 29 atoms in the third layer (not in the
ideal 4$\times$7 arrngement) and 28 in the fourth. This suggests that
the period along $y$ in the thermodynamic limit be somewhat
larger than the value $b/7$ allowed by the simulation cell.
Finally, Fig.~(\ref{esagoni}) also shows in the left panel a stripe along 
$x$ at low coverage in the fourth layer, consistent with Fig.~(10) 
in the main text.

\subsection{Adsorption on the lower surface}
\begin{figure}[h!]
\includegraphics[width=13cm]{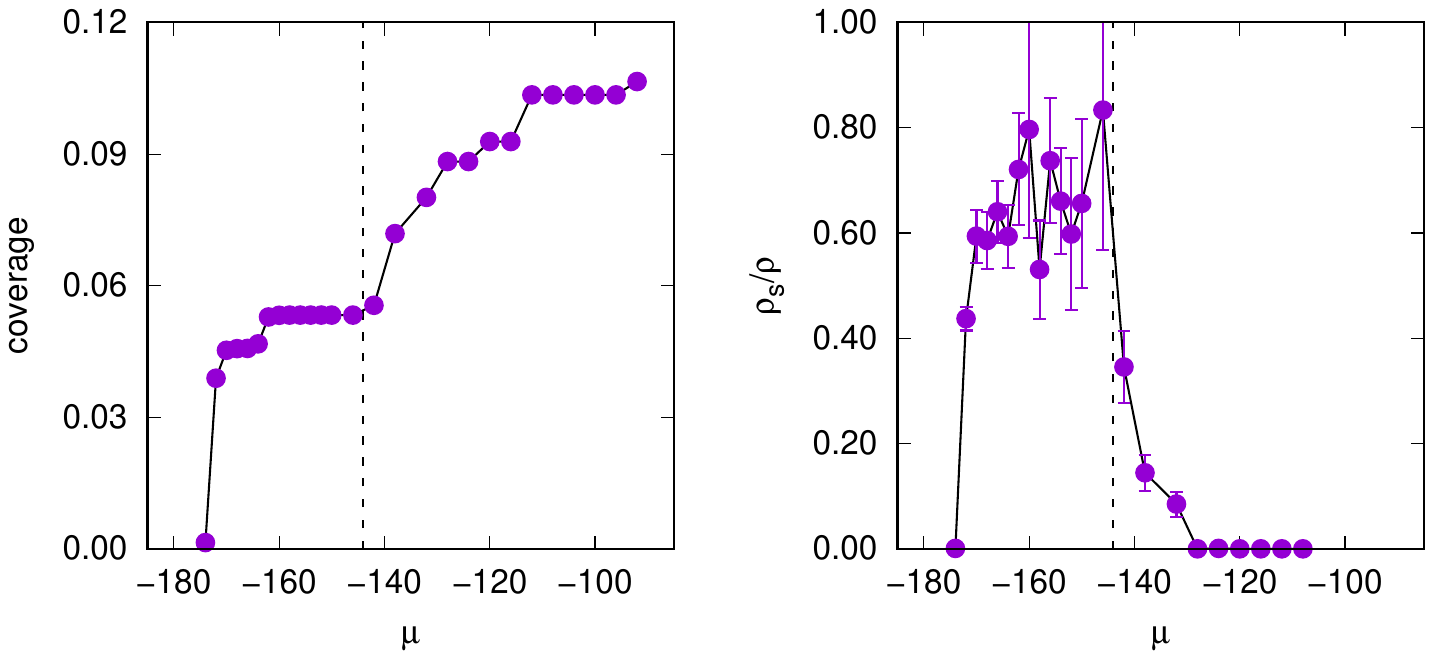}
\caption{Coverage (left) and superfluid fraction (right) of $^4$He
adsorbed on the lower surface of BPH-rect as a function of the chemical
potential at $T=0.5$~K. The dashed lines indicate promotion to the
second layer.}
\label{fig_17}
\end{figure}

\begin{figure}[h!]
\includegraphics[width=10cm]{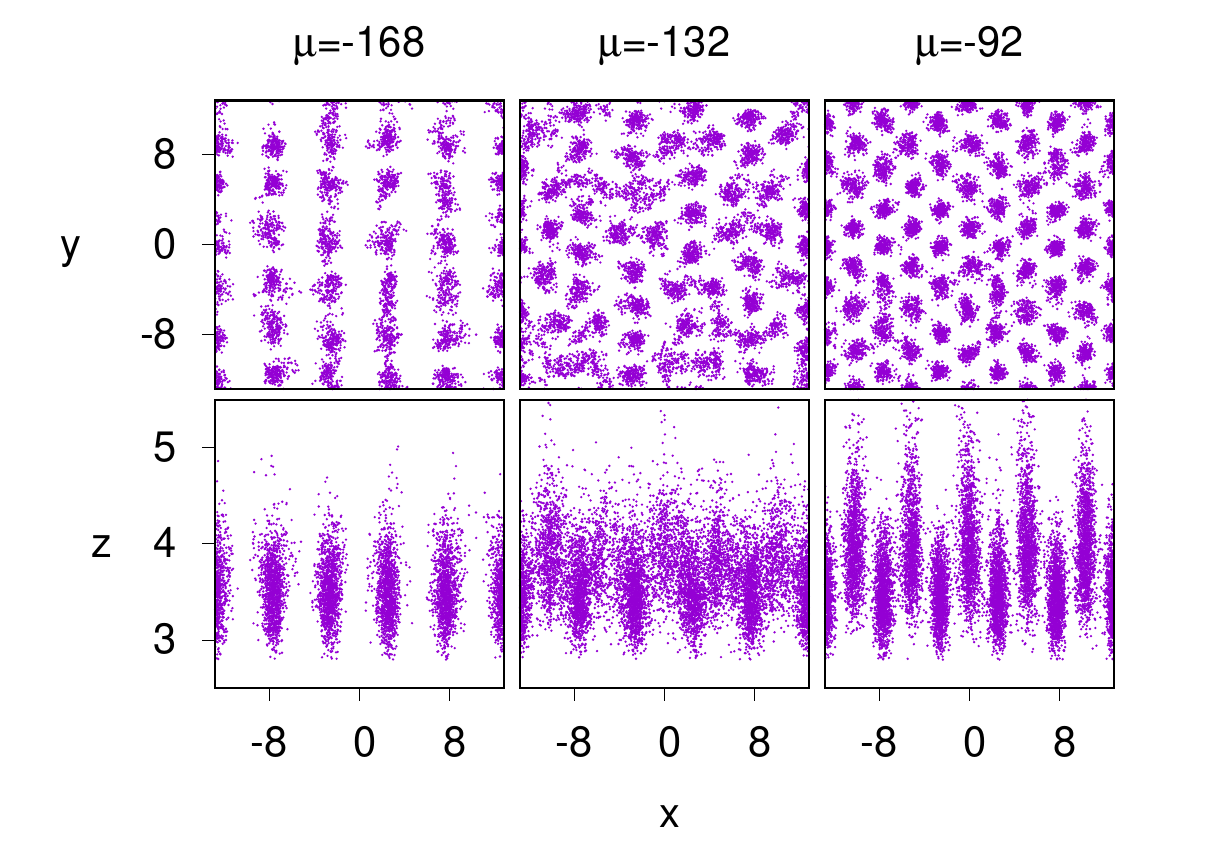}
\caption{
Snapshots of $^4$He atoms adsorbed on the lower surface of BPH-rect for various coverages.
Upper panels are top views, lower panels are $(x,z)$ side views.
}
\label{sotto}
\end{figure}

\begin{figure}[h!]
\includegraphics[width=\columnwidth]{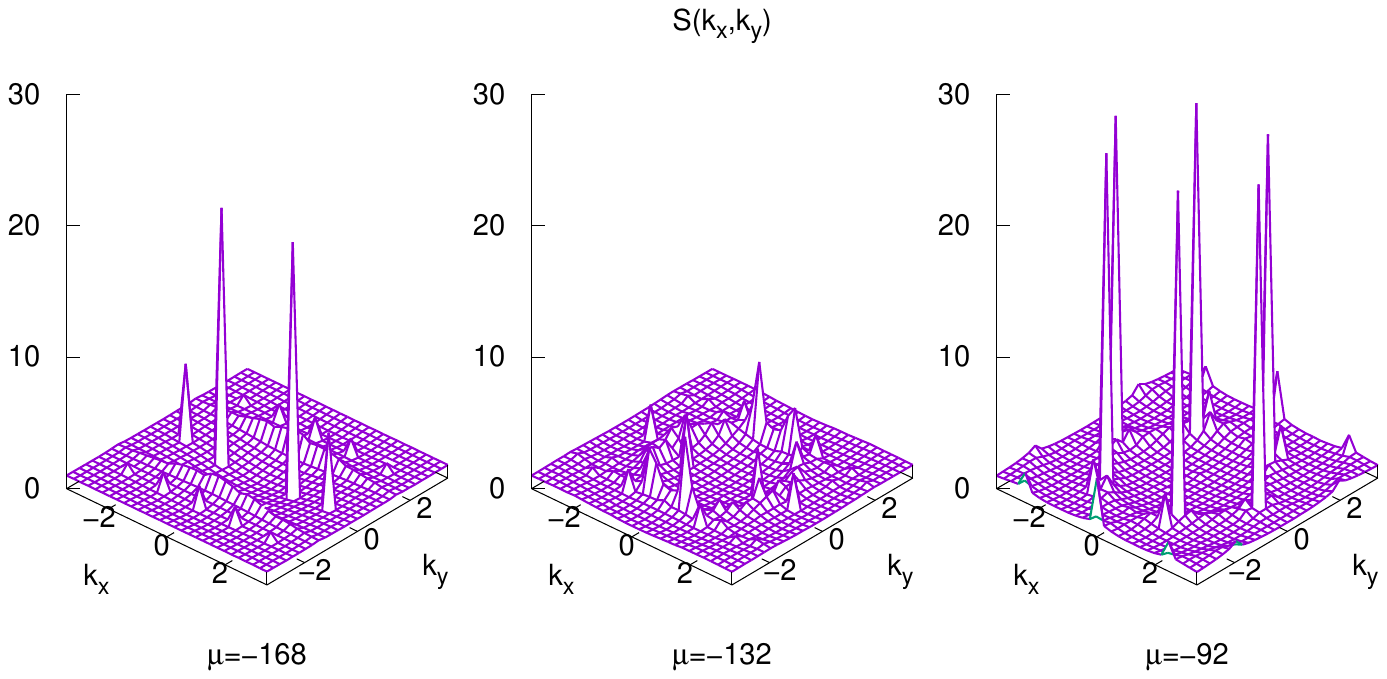}
\caption{
Two-dimensional structure factor for various coverages
of the lower surface of BPH-rect.
}
\label{sk_new}
\end{figure}

The substrate-He potential for the lower surface of BPH-rect has smaller 
barriers
between adsorption sites (see Table II in the main text). The $^4$He 
atoms in the first layer are not pinned one-on-one to the sites
and sustain superfluid flow.
Superfluidity persists up to slightly beyond completion of the first layer
(see Fig.~\ref{fig_17}). As the coverage of the second layer increases,
a buckled triangular crystal develops, incommensurate to the substrate.
This is illustrated by snapshots of the path integral configurations
in Fig.~\ref{sotto} and by the structure factor $S(k_x,k_y)$ in 
Fig.~\ref{sk_new}.

\subsection{Results with the DFT-D2 potential}

\begin{figure}[h!]
\includegraphics[width=\columnwidth]{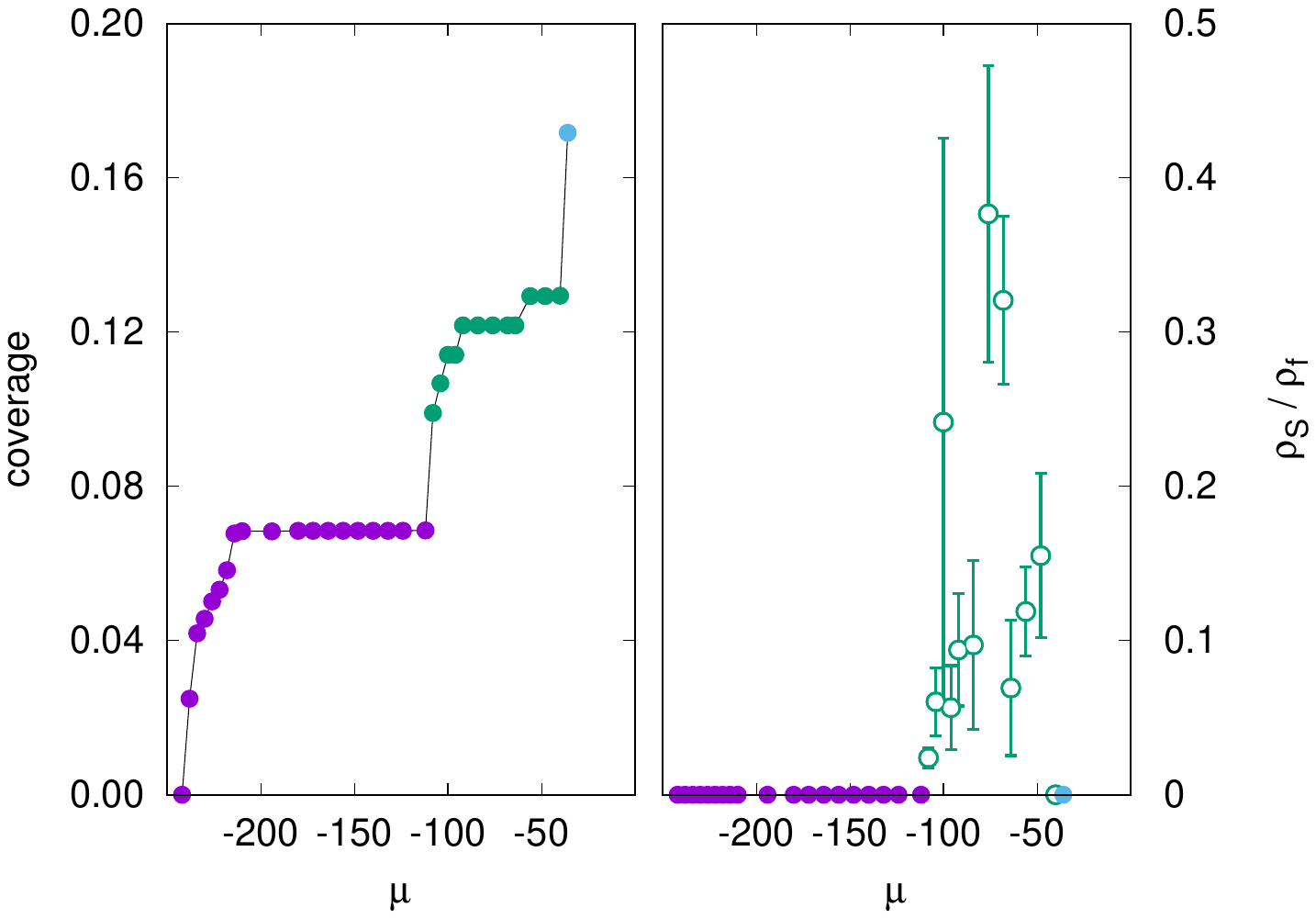}
\caption{
Coverage (left panel) and superfluid fraction (right panel)
of $^4$He adsorbed on BPH-rect as a function of the chemical potential at 
$T=0.5$, calculated with the DFT-D2 potential. 
Different colors correspond to different layers. In the right
panel open circles indicate the superfluid fraction along the $y$ direction.
}
\label{np_a}
\end{figure}

\begin{figure}[h!]
\includegraphics[width=\columnwidth]{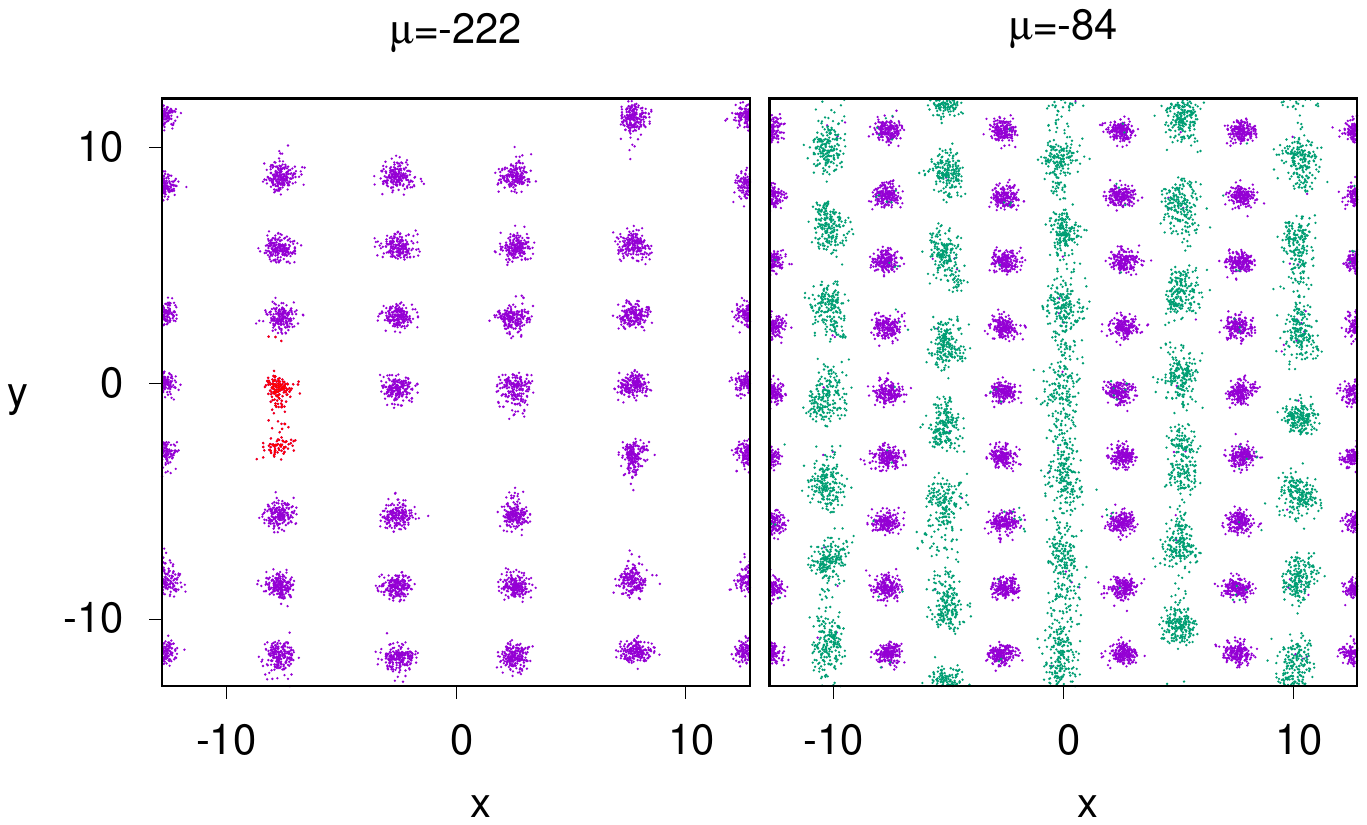}
\caption{Snapshots from simulations of $^4$He adsorbed on BPH-rect at $T=0.5$
using the DFT-D2 potential.
Left panel: 35 $^4$He atoms partially filling the
45 adsorption sites present in the simulation cell. Each particle is represented
by a sequence of beads, shown here with a stride of five. Red beads denote 
particles hopping between two sites.
Right panel: 80 $^4$ He atoms distributed in the first two adsorption layers.
Different colors correspond to different layers.
}
\label{coord_a}
\end{figure}

This Section presents results for the upper surface of BPH-rect calculated
with the DFT-D2 potential, which gives more negative binding energies
than the rVV10 potential and somewhat stronger corrugation (see Table I in
the main text). Correspondingly, the adsorbed $^4$He atoms are somewhat
more localized. However, the main features of the first two adsorption 
layer obtained with the rVV10 potential are preserved: (i) in the first
layer we have a gas of atoms that hop between adsorption sites
along the $y$ direction (with no superfluidity, at the temperature studied), 
until all sites are occupied, and (ii) the second layer is a fluid with up to
8 atoms per channel, with a non-zero superfluid density that persists
until promotion to the third layer. This is illustrated in Figs.~\ref{np_a}
and \ref{coord_a}, to be directly compared with the corresponding
rVV10 results in Fig. 8 in the main text, Fig. 1 in this SM, and the top
panel at $\mu=-92$ of Fig. 9 in the main text.

For higher coverages the effect of the
adsorption potential is increasingly screened by the 
underlying $^4$He atoms, and we expect the difference between
rVV10 and DFT-D2 potentials to become even less relevant.


\begin{thebibliography}{99}
\bibitem{dft-d2} We have verified that the DFT-D2 potential gives 
qualitatively similar results.
\end{thebibliography}
\end{document}